\documentclass[aps,pre,showpacs]{revtex4-2}

\usepackage{bm}
\usepackage{amsmath}
 \usepackage{amssymb}
 \usepackage{graphicx}
\usepackage{color}

\begin{document}

\title{Pulse Instabilities Can Shape Virus-Immune Co-evolution}
\author{David A. Kessler$^*$ }
\affiliation{Department of Physics, Bar-Ilan University, Ramat-Gan 52900, Israel}

\author{Herbert Levine }
\affiliation{Center for Theoretical Biological Physics and Depts. of Physics and Bioengineering, Northeastern University, Boston, MA 02215, U.S.A.}
\date{\today}

\begin{abstract}
Adaptive immune systems engage in an arms race with evolving viruses, trying to generate new responses to viral strains that continually move away from the set of genetically-varying strains that have already elicited a functional immune response. It has been argued that this dynamical process can lead to a propagating pulse of an ever-changing viral population and concomitant immune response. Here, we introduce a new stochastic model of viral-host co-evolution, taking into account finite-sized host populations and varying processes of immune "forgetting".  Using both stochastic and determinstic calculations, we show that there is indeed a possible pulse solution, but for a large host population size and for finite memory capacity, the pulse becomes unstable to the generation of new infections in its wake. This instability leads to an extended endemic infection pattern, demonstrating that the population-level behavior of virus infections can exhibit a wider range of behavior than had been previously realized . 
\end{abstract}

\maketitle

\section{Introduction} 
Viruses form a convenient forum for studying Darwinian evolution~\cite{review}. They reproduce quickly, giving rise to large populations, and are therefore more deterministic in their dynamics than populations of macroscopic living creatures~\cite{novella}. Although their reproduction is asexual, nevertheless they can even exhibit rudimentary forms of genetic recombination when different viral strains with multi-compartmented genomes invade a common cell~\cite{recombination}. 

A powerful conceptual approach for studying viral evolutionary dynamics involves projecting the dynamics onto a low-dimensional fitness space~\cite{kepler,levine-prl,rouzine}. This approach led to the discovery of the possibility of a propagating solitary wave of viral strains on a linear fitness landscape, propelled by the more rapid growth of the advancing edge as compared to the out-competed variants at the decaying tail. Interestingly, determining the velocity of this wave turned out to be a challenging statistical physics problem, as the naive continuum PDE governing the viral fitness density exhibits a finite time singularity~\cite{levine-prl} and hence necessitates a more nuanced treatment of this limit~\cite{rouzine,desai-fisher}. More specifically, a modified PDE approach~\cite{levine-prl,rouzine,krug} led to the prediction of logarithmic scaling of the advancing speed with viral population size, a result in agreement with simulations of stochastic versions of the model and with the exact result for a specific model variant~\cite{halla}.

Most recently, a series of papers~\cite{rouzine2,walczak1,walczak2} have generalized this framework to study the co-evolution of the viral population with the immune system. Here, individual hosts generate immune responses that target the virus even as the virus tries to mutate away from immune recognition. Of particular importance in this class of models is the extent to which the immune response, mediated for example via antibody recognition of the viral particle, decays as a function of mutational distance~\cite{japanese}. One possible simple assumption is that recognition falls off as an exponential of the Hamming distance between D-dimensional viral genomes;  one can also imagine non-symmetric fall-offs ~\cite{rouzine2}, but the reasoning behind this assumption is not obvious. As pointed out in these papers, the immune response to previous infections can create an effective gradient in viral fitness; the further away from the bulk of the population, the weaker the immune suppression and hence the faster the growth. This means that the propagation of a single pulse along some direction in this D-dimensional space is in the same class~\cite{cohen-levine,neher} as the aforementioned pure viral evolution problem. The fact that such a pulse solution is localized in the direction transverse to the propagation ~\cite{walczak1} makes this type of one-dimensional solution semi-quantitatively relevant even for an arbitrary number of phenotypic dimensions.  Thus, predictions from this modeling framework can be very informative regarding general patterns of viral dynamics.

In this paper, we focus our attention on the one-dimensional case and study a generalized version of the model of Ref.~\cite{walczak1}. Our study considers the stochastic dynamics together with the modified PDE approach wherein the leading effects of stochasticity (here, demographic noise) are taken into account by a small-population cutoff of the interaction dynamics. In the deterministic version, we show that a stable single pulse solution exists for smaller host populations but that it becomes (nonlinearly) unstable above some critical population size. This instability leads, immediately beyond the critical size, to a patterned tail and eventually, at even larger sizes, to a featureless tail.  Both of these are forms of endemic infections. These expanding tails cause the front to slow down with time as they exhaust the number of available hosts for new variants to infect.  Stochastic dynamics blurs the distinction between the patterned and unpatterned tail, but otherwise exhibits a similar transition from a stable pulse to an endemic state exhibiting a slowly widening and eventually stationary band of active variants. Thus, the long-time behavior is often a non-spreading statistically steady infection pattern. This new type of behavior for viral-host co-evolutionary dynamics,  possibly relevant both for human respiratory diseases and for the infection of bacteria by phages, is the major result of this paper.

\section{Stochastic model}
\begin{figure}[t]
\includegraphics[width=.8\linewidth]{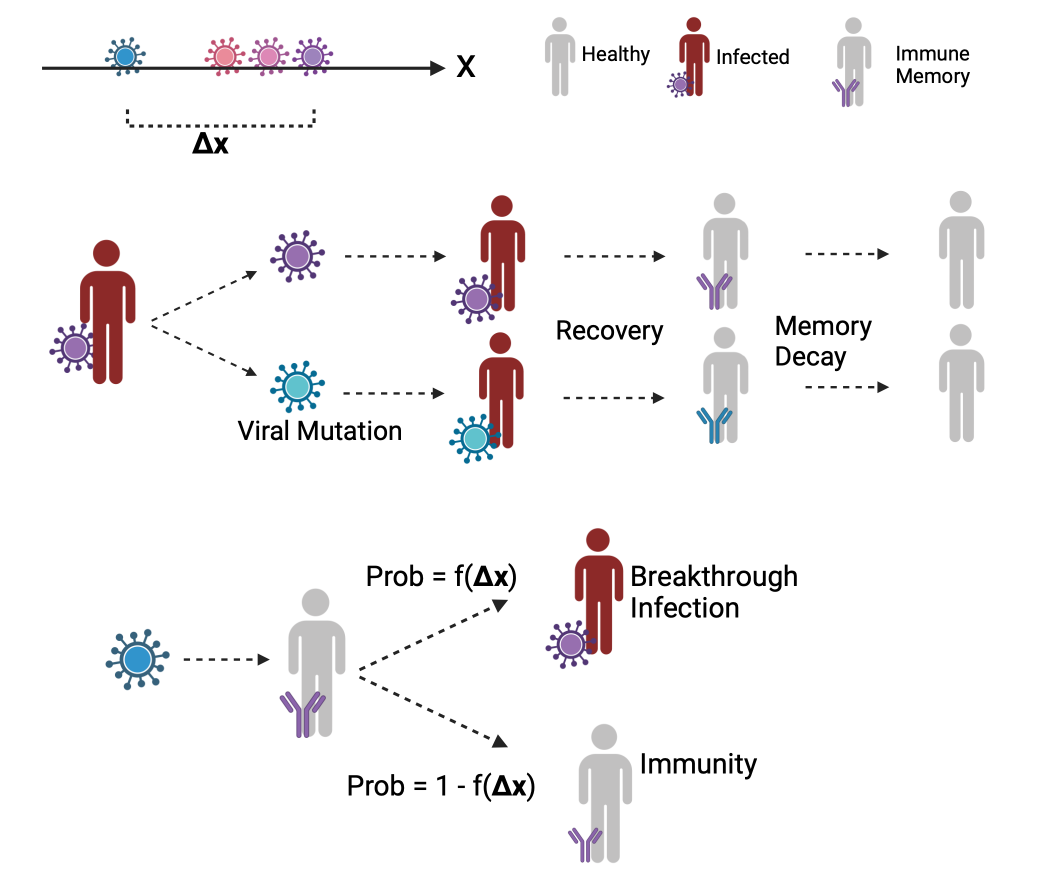}
\caption{  A schematic view of the processes considered in our model. The space of allowed virions is taken to be one-dimensional, labeled by coordinate $x$. There is a release of virions by an infected person, some of which can be mutated by moving one unit in $x$. These virions then go on to infect other people, depending on the state of their immunity. Upon recovery, a previously infected person stores a memory of the infecting virus. Memories can be lost in a variety of ways, as discussed in the text. Finally, immune protection against one virus by the memory of another wanes as the distance between them increases. \label{schematic}}
\end{figure}
\subsection{Formulation}
The basic process we propose to study involves an evolving virus that infects individual hosts in a population. Hosts respond to infection by launching an immune response, best thought of as selecting antibodies that effectively neutralize viral particles emitted by infected individuals. Of specific interest is the fact that such antibodies will also, to some extent, neutralize mutant strains. This extent depends on exactly how far the mutant has strayed from the original.

The actual problem of determining the expected falloff of immune effectiveness is rather complex and undoubtedly context-dependent. However, interesting insights into the behavior of viral immune co-evolution have been obtained by making simplified assumptions about this feature of ``shape space”~\cite{perelson}.
The simplest assumption is that there is a monotonic mapping of immune effectiveness fall-off as a function of how many relevant mutations have been accumulated by a particular viral strain~\cite{walczak1, japanese}. Using this approach, we can define
a cross-reactivity function $ g(|\vec{x}-\vec{x}'|)$, 
where the fall-off function $g$ is the ratio of effectiveness at a distance  $d=|\vec{x}-\vec{x}'|)$ to that at zero distance; $g$ obviously approaches unity at small distances and vanishes at large distances. This assumption is reminiscent of that used in ecological models of competing species~\cite{pigolotti,rogers}, where niche selectivity is assumed to give rise to a non-local but decaying interaction between specialists optimized to grow in specific environments. In what follows, we will assume that $g$ takes the form of an exponential of width $w$.  As noted above, we here consider a discrete one-dimensional shape-space, with position (i.e., strain) labeled by the index $i$.

Our model assumes that there are a total of $N_h$ individual hosts, each of which can be infected by at most one viral strain at a time. The number of infected hosts is $N_I (t)$, which obviously cannot exceed $N_h$. The possible events at some given moment are the recovery of one of the infected hosts (at total rate $rN_I$) or the release by one of the hosts of a new virion at rate  $rN_I R_0$. Here the base reproduction rate $R_0$ is the mean total number of ``effective" virions, namely virions that are directly able to infect new individuals, released by a single infected host during its duration of being infected before recovery. Given the total rate of events, we choose the time step according to the standard Gillespie algorithm and then pick which event occurs with the relative probabilities $1/(1+R_0)$ for clearance and $R_0/(1+R_0)$ for virion release, and for both event types the host in question is chosen at random from the infected pool. Clearance is effected by removing the chosen host from the infected list, reducing $N_I$ by one, and updating its immune memory (see below). The release of a virion is accompanied by its attempt to infect a new host. We allow this released virion to be slightly modified from the strain that caused the infection in question; specifically, we allow its index $i$ in shape-space to mutate either up or down by one with equal probabilities given by $\mu /2$.  If the randomly chosen new host is currently uninfected, then infection will occur with a probability $p_\textit{inf}$ that depends on the degree to which that host has immunity to that particular virion strain. 

The model is completed by describing how the immune state of an individual host is updated and how that state determines the infection probability. We will assume that the host carries multiple immune ``memories", corresponding to its previous infection history. The immune memory is realized in the matrix $\textit{mem} (j,i)$, storing the memory of host $j$ having been infected by viral strain $i$. There is no rule that memories need to be distinct, as hosts can have been multiply infected in the past by the same strain, with each exposure bolstering the cell's immunity to that strain. In our update scheme, every time a host recovers, it stores a memory of the strain from which it has just recovered. In the usual case where the number of existing memories for that host is equal to the memory capacity, this new memory overwrites a randomly chosen previous memory. Finally, we have the explicit formula for the probability of infecting host $j$ by viral strain $k$
\begin{equation}
p_\textit{inf} (j,k) \ =  \prod _{i=1}^{M} \left( 1- p_{0} g( |mem(j,i) - k|) \right)
\end{equation}
Finally, we choose $p_0 = (1- p_{00} ^{1/M} )$, which therefore gives a bare infection rate of $p_{00}$ for any hosts all of whose memories are equal to that of the strain with which it is being challenged. 

There are obviously many parameters in this model and it would be impossible to exhaustively vary all of them. As our interest here is in revealing the types of behavior that can occur, we will initially stick to the choices $p_{00}=0.015$, $M=6$, giving $p_0\approx 0.50$. We will consider two values for the mutation rate, a high value ($0.5$) which creates relatively smooth profiles and hence allows comparison to our deterministic PDE approach, and a perhaps more biologically reasonable value ($0.025$).  In Section \ref{sec:memory}, we turn to a closer investigation of the impact of the number of memories, and the details of the memory loss mechanism, on the dynamics.

\begin{table}
\begin{center}
    \begin{tabular}{|l|l|}
      \hline
      \ \ Parameter/Symbol \ \  & \ \ Meaning (Default Value)\ \  \\\hline
      $x$ & Viral strain\\\hline
      $n(x)$ & Number of Infecteds with strain $x$ \\\hline
    $I(x)$ & Production rate of new infecteds with variant x \\\hline
      $V(x)$ & Rate of emission of virion strain $x$ 
      \\\hline
      $E(x)$ & Rate of virion emission by infecteds with strain $x$ \\\hline
      $f_\epsilon(n(x))$ & Cutoff function preventing infection  when the number of infecteds when the local number of infecteds is too low\\\hline
      $p_\textit{inf}(j,x)$ & The infection probability for  host $j$ exposed to
      strain $x$  becoming infected 
      \\\hline
      $g(\Delta x)$ & The function governing the cross-immunity between strains at a distance $\Delta x$
      \\\hline
    $\textit{mem}(k,i)$ & The $i$th memory of host $k$
    \\\hline
    $p_{inf}(x)$ & The probability that a random host will become infected if exposed to strain $x$
    \\\hline
    $\rho_m(x)$ & The population density of memories of strain $x$
    \\\hline
    $Q(x)$ & An auxiliary field used to compute $p_\textit{inf}(x)$ from $\rho_m$
    \\\hline
    $N_I$ & The total number of infecteds\\\hline
    $N_h$ & The total number of hosts\\\hline
    $R_0$ & The expected number of exposed hosts directly due to a single infection (3.8)
    \\\hline
    $\mu$ & Probability of an emitted virion being a different strain than its parent ($0.025$, $0.5$)
    \\\hline
    $M$ & Number of memories per host (6)
    \\\hline
    $w$ & Width of cross-immunity  \\ \hline
    $\epsilon$ & Cutoff local density below which infections do not occur
    \\\hline
    $\alpha$ & Exponent entering in the cutoff function $f_\epsilon$ (4)
    \\\hline
    $p_0$ & A parameter quantifying the degree of partial immunity for the initial infecting strain (0.5)
    \\\hline
    \end{tabular}
\end{center}
\caption{}
 \end{table}

\begin{figure}[h]
\includegraphics[width=.32\linewidth,height=.22\linewidth]{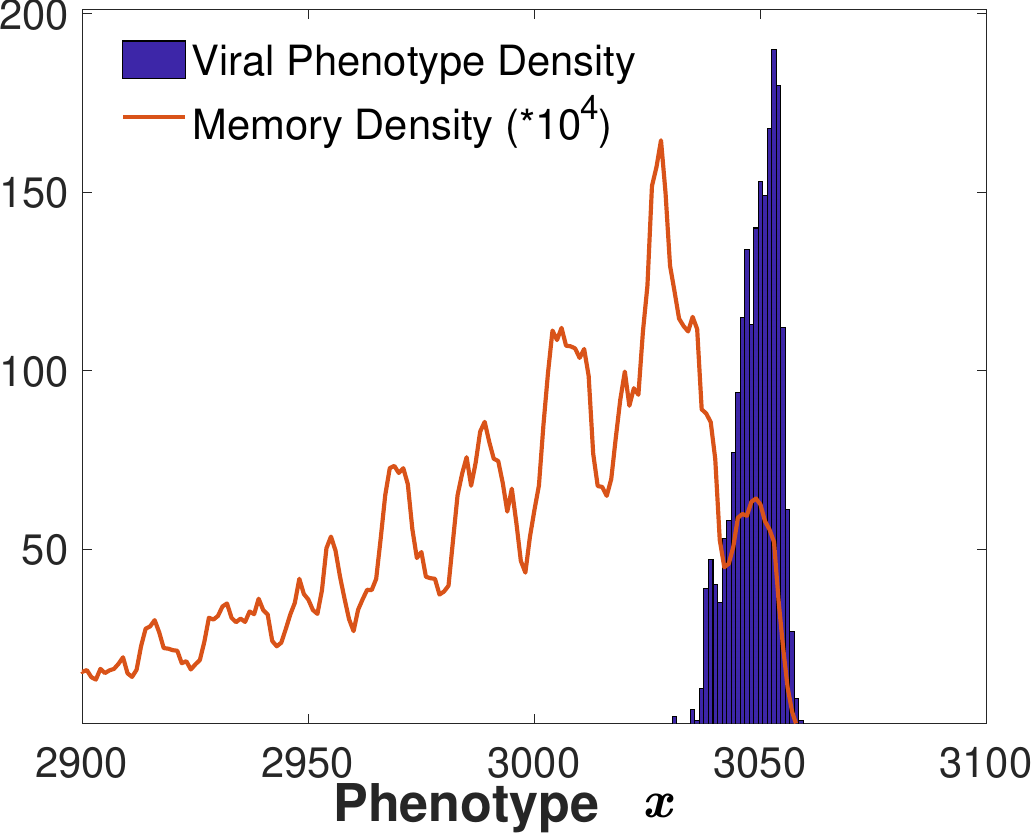}
\includegraphics[width=.3\linewidth]{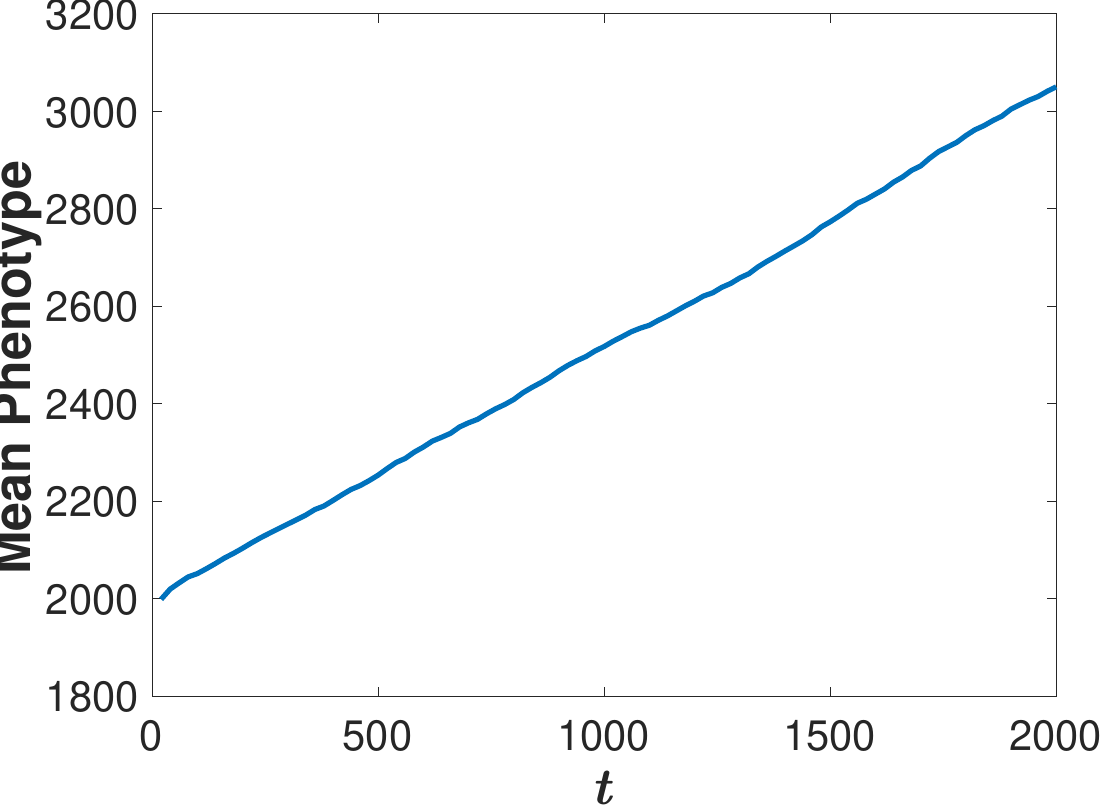}
\includegraphics[width=.3\linewidth]{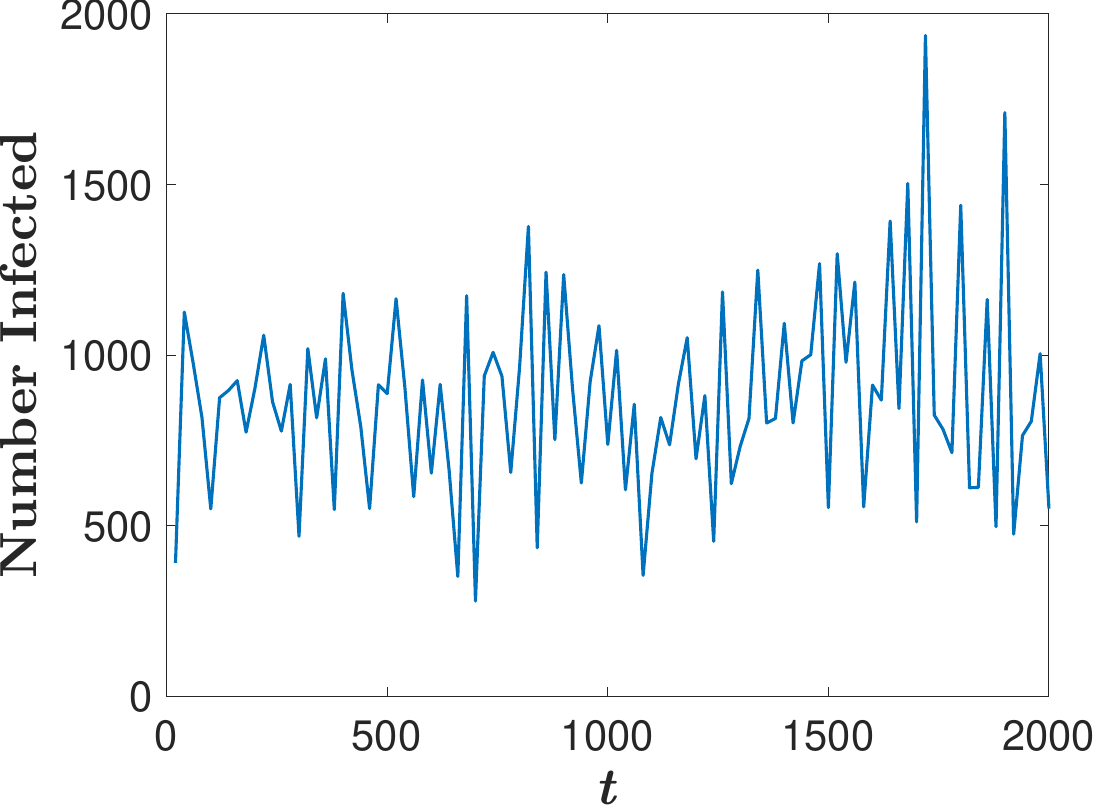}
%
\caption{A stochastic traveling pulse, for $w=40$, $\mu=0.5$, $N_h=2\cdot 10^4$. Left panel:  An instantaneous snapshot of the viral strain denstiy profile
$n(x,t)$ and the memory density $\rho_m(x)$; Center panel:  The mean of the positions (phenotypes) of the infected individuals, as a function of time $t$. Right panel:  The total number of infected individuals, $N_I(t)$, as a function of time. \label{fig_1}}
\end{figure}

\subsection{Pulse solutions}
Let us start with a simulation with $w=40$, $\mu = 0.5$ and $N_h = 2\cdot10^4$.  We assume that initially some number of individuals are infected by a single viral strain and there is a completely blank slate of memories for each of the hosts.  Starting with 40 initial infecteds, the time evolution of the system shows three distinct possibilities. One possibility is that the infected population collapses almost immediately; this happened for four out of 50 independent simulations. This behavior occurs because viral phenotypic diffusion, as characterized by $\mu$, does not occur fast enough to overcome immune suppression. The percentage of immediate failures can be reduced by starting from a more spread-out initial state.

The other two behaviors are more interesting. One type is a relatively stable single propagating pulse~\cite{levine-prl,walczak1}; this state was reached in approximately 60\% of the simulations, but again this number can depend on the detailed initial conditions. A snapshot of this behavior is shown in Fig. \ref{fig_1}a along with the time history of the mean infection position along the phenotypic axis (Fig. \ref{fig_1}b) and the history of $N_I (t)$ (Fig. \ref{fig_1}c).  For this parameter set, the pulse, once formed, is very stable, lasting at least for many tens of thousands of simulation steps. This stability is due to the fact that the range of fluctuation as exhibited in Fig. \ref{fig_1}c keeps the population number quite far from the absorbing state at $N_I=0$. Thus, while the fluctuations are indeed relatively large, the qualitative behavior seen here is similar to what would be seen in a deterministic model exhibited for example, in the work of Ref.~\cite{walczak1}. Note that the average value of $N_I$ is much smaller than $N_h$ and so, for this state, model variants such as the cited ones that neglect the finite size host pool may be reasonable simplifications of our more complete formulation. 

\begin{figure}[b]
\includegraphics[width=.48\linewidth]{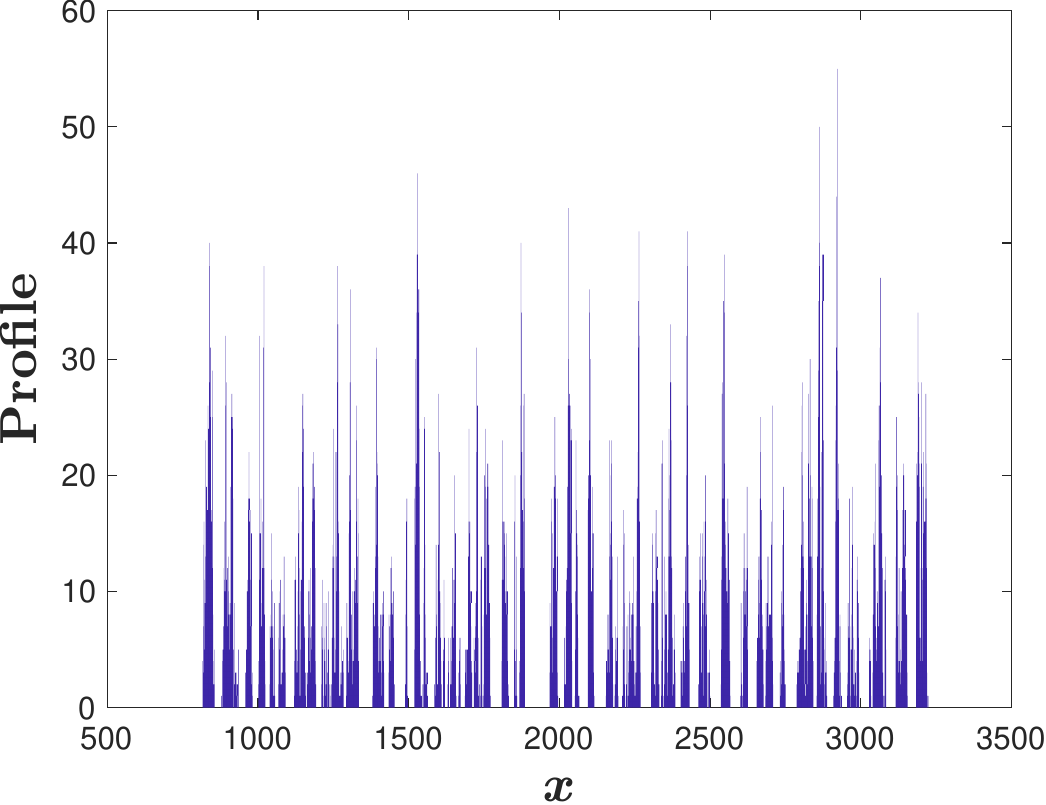}
\includegraphics[width=.48\linewidth]{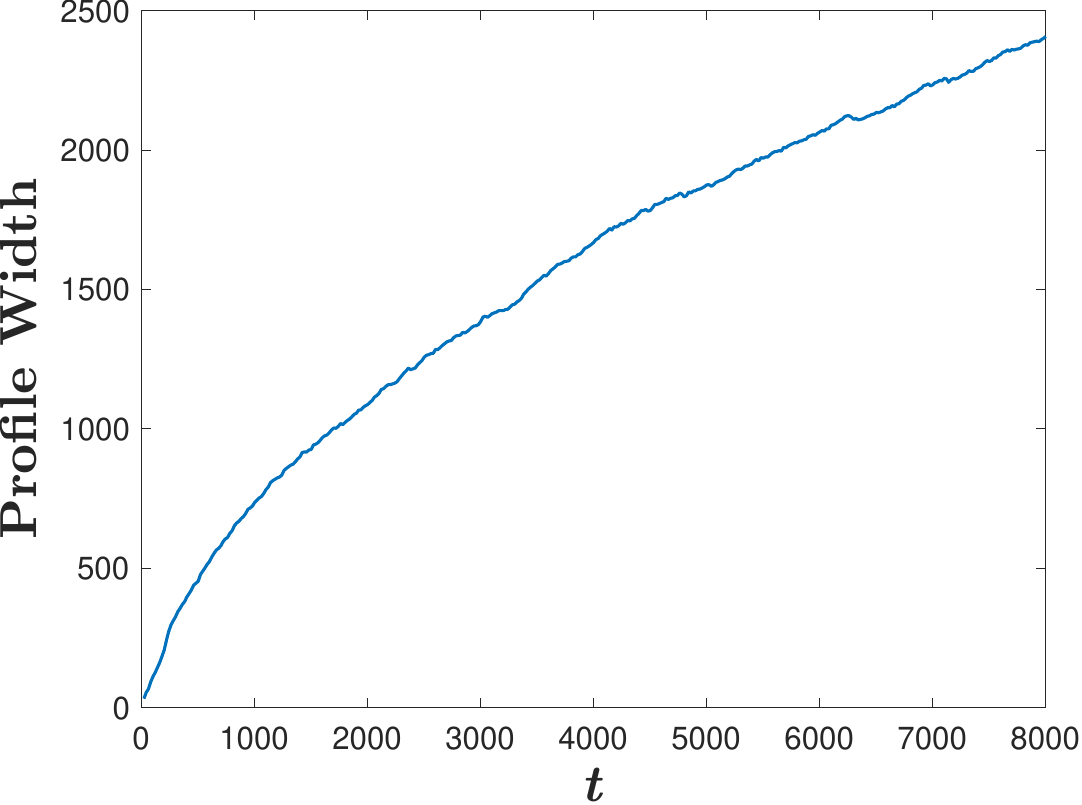}
%
\caption{The mound pattern for $w=40$, $\mu = 0.5$ and $N_h = 2\cdot10^4$. Left panel: An instantaneous snapshot of the viral phenotype densities $n(x,t)$, showing an extended tail in the wake of an originally propagating pulse. Right panel:  The mound width, namely the distance between the largest and small positions of individuals in the mound, as a function of $t$. \label{fig_2}}
\end{figure}

The second type of solution arises when the wake of the pulse becomes unstable to the emission of a new population peak. This leads eventually to a broadly distributed infection pattern, as seen for example in Fig. \ref{fig_2}a. The overall width of the pattern increases very slowly (Fig. \ref{fig_2}b) until eventually saturating, even as the infected population number remains relatively constant. As $N_h$ is increased, an increasing percentage of runs result in this second type of dynamical behavior. We will return shortly to a quantitative characterization of these solutions.

\begin{figure}[b]
\includegraphics[width=.48\linewidth]{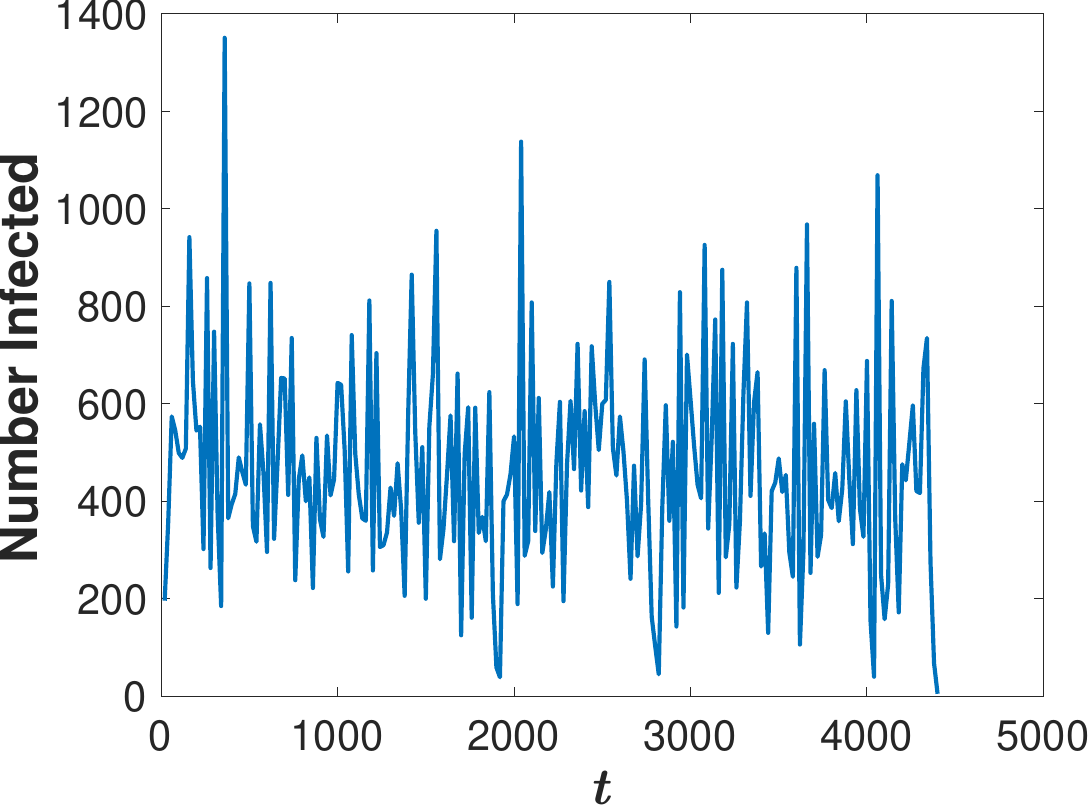}
\includegraphics[width=.48\linewidth]{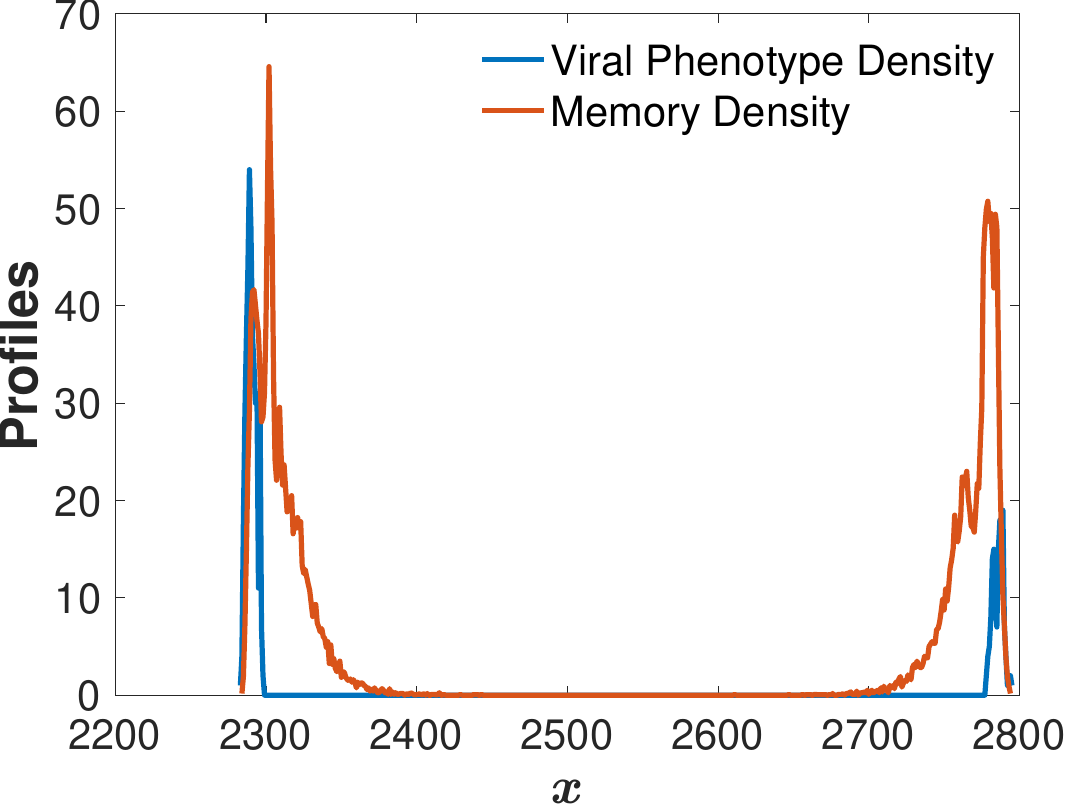}
%
\caption{Left panel: The total number of infected hosts, $N_I(t)$, showing the collapse of a traveling pulse for $N_h=1.2\cdot 10^4$. Right panel: Two counter-propagating pulses for $N_h=5000$, showing the viral phenotype profile $n(x)$ in black and the (unnormalized) memory profile $\rho_M (k)  \equiv \sum_{i,j} \delta_{mem(i,j),k}$ in red. For both panels, $w=40$ and $\mu=0.5$. \label{fig_3}}
\end{figure}

What happens to the pulse solution if $N_h$ is lowered? Carrying out simulations at $N_h = 12,000$ reveals the fact that the pulse solution is in fact metastable. In Fig. \ref{fig_3}a we plot $N_I(t)$ for a pulse that spontaneously collapsed after a significant amount of time. Clearly, going to a lower host population size has lowered the mean value of $N_I$ for the pulse, thereby putting the absorbing state within the range of large but not excessively large fluctuations. The population depicted in the figure survived close calls at $t \sim 1900$, $2800$, and $4000$ before finally succumbing at $t \sim 4400$. Although we did not carry out an exhaustive analysis, our data is consistent with a constant rate of decay of the probability of survival of the pulse, which of course implies an exponential distribution of survival times. The decay rate appears to be a strongly decreasing function of $N_h$ accounting for the lack of observations of collapse for our simulations at $N_h=20K$. 

If a single pulse solution exists and is linearly stable, one can ask about multi-pulse solutions with either counter-propagating or co-propagating infected host populations that are far enough apart to minimize cross-reactivity. Fig 3b shows an example of a counter-propagating pair, at the much lower value of $N_h=5000$. Interestingly, this pattern is not typically selected from generic localized initial conditions,  unlike what is observed in excitable media with purely local inhibition where each edge of the initial data generates an independent and oppositely propagating pulse. Furthermore, no matter how far apart the pulses are, they continue to interact via the limitations on immune memory; a host becoming infected by a far-away virion will degrade its immune response to the near virion by which it was most recently infected.  In other words, the model has long-range coupling in phenotypic space. This tends to diminish the ability of the pulse wake to remain stable and thereby lowers the range of the overall host population size for which the two-pulse solution is stable.

We also carried out calculations at values of the parameters which may be more physiological and which more strongly emphasize stochasticity. Specifically, we set $w=10$ which makes the immune shadow of an infection pattern much less spatially extended. To partially compensate, we set  $\mu =  .025$; note that this is close to the naive estimate that one can change these parameters while maintaining the dimensionless ratio $\tilde{\mu} = \mu/w^2$  constant and expect only a limited change of behavior. Essentially, this modified system exhibits behavior that is similar to that of the previous parameter set. In particular,  most but not all simulations converge quickly to a pulse solution, which subsequently dies off due to fluctuations at a constant rate. We will discuss in the next section the other generic behavior mentioned above, a propagating pulse that spontaneously nucleates new virion peaks in its wake, leading to a phenotypically extended pattern

\subsection{Extended states}

\begin{figure}[b]
\includegraphics[width=.32\linewidth]{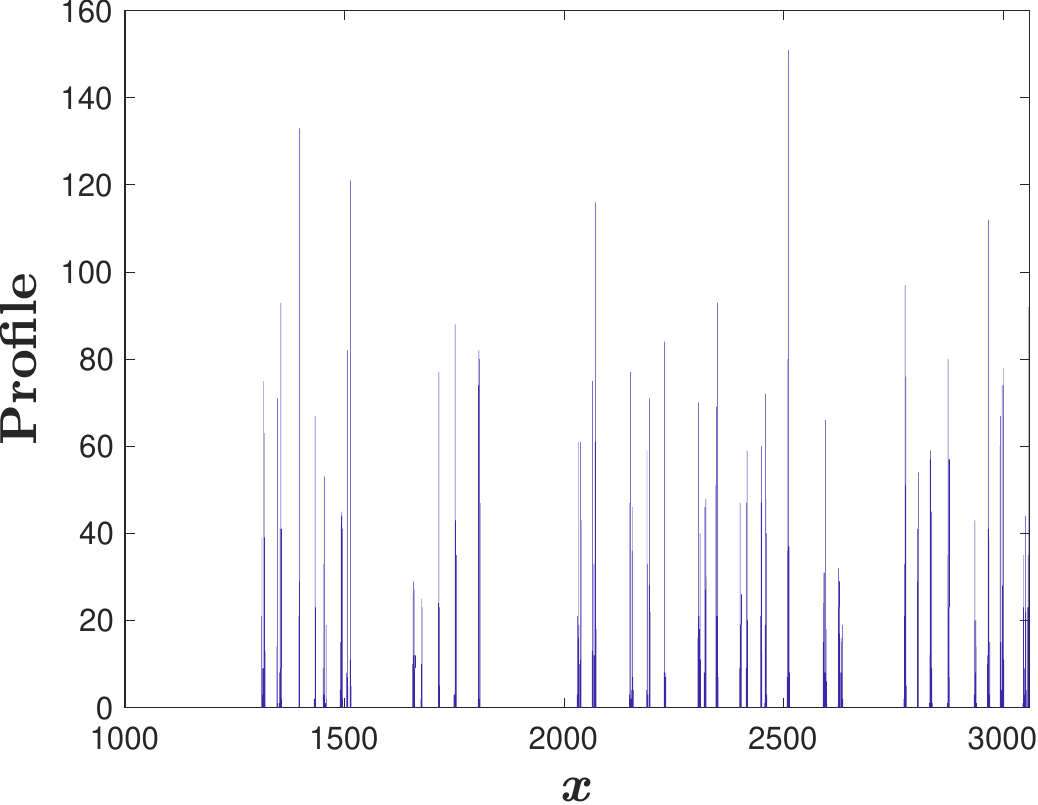}
\includegraphics[width=.32\linewidth]{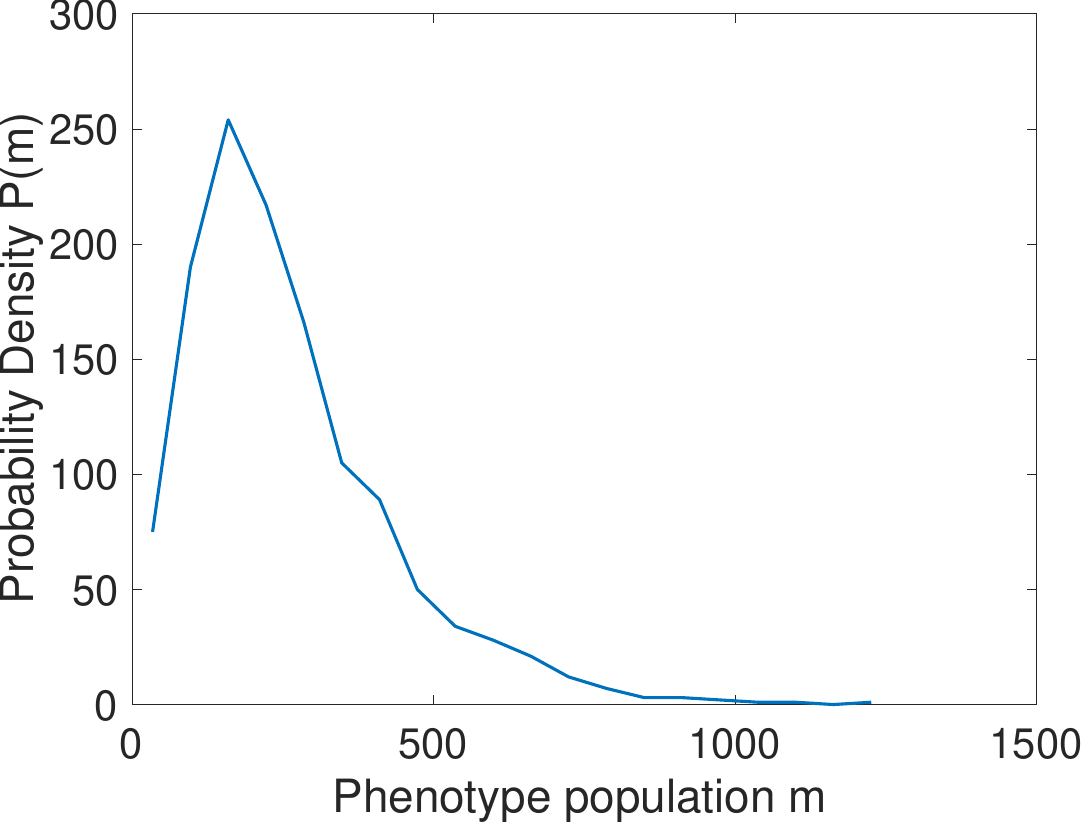}
\includegraphics[width=.32\linewidth]{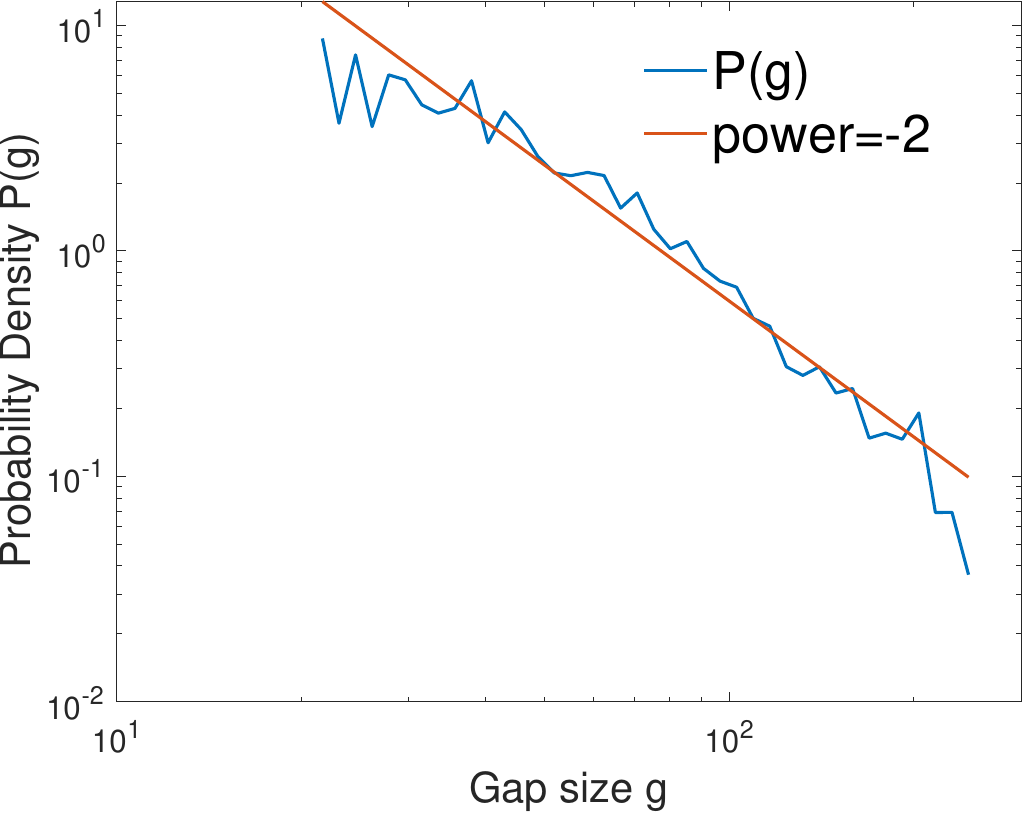}
\caption{Stochastic simulation for $w=10$, $\mu=0.025$, $N_h=10^4$. Left panel: Infection profile $n(x)$.  Center panel: Distribution of effective phenotype (defined as a set of strains without a gap of $2w$ separating them) population sizes $P(m)$.  Right panel: Distribution of gap sizes, $P(g)$, between effective phenotypes, showing an approximate $P(g)\sim g^{-2}$ power law. \label{fig_4}}
\end{figure}

As already mentioned, simulations at large $N_h$ typically lead to extended patterns, with localized patches of infection separated by gaps, a range of phenotypes not present in the population. As has already been seen in Fig. \ref{fig_2}b, the width of the overall structure increases slowly over time and takes a long time to reach a very noisy steady state. Since the overall population remains roughly constant (at approximately $2/3 N_h$), this implies that either the patch size must be decreasing or the typical gap must be widening during this long transient. Now, just as we saw for isolated pulses, patches cannot remain stable for long times when their size becomes too small. Hence, as patches shrink, they will disappear and lead to larger gaps between neighboring patches. This is balanced initially by the creation of new patches in the wake of existing ones (see below), but this effect diminishes as the number of patches becomes large and $N_I$ reaches its steady-state value much above that of a single pulse. Also, patches have a small repulsive interaction due to residual immune memory effects, All this means that over the transient period, the major effect is a spreading apart of the patches.  This will only cease when the rate at which the outermost patches disappear matches the average rate at which they move outward. This behavior leads to a very diffuse pattern in the very long time limit.

In Fig. \ref{fig_4}a, we present a second example of this structure, at $N_h =10^4$, $w= 10$ and $\mu = 0.025$, now focusing on simulation times long enough to have reached the very noisy steady state. As advertised, the basic structure is that of relatively isolated patches of infected host density separated by gaps that are much larger than $w$. In Figs. \ref{fig_4}b and \ref{fig_4}c we plot respectively the distribution of patch populations and of the gaps. The patch population size exhibits a peaked distribution and hence the patches can be characterized by a typical population size. On the other hand, the gap distribution is power-law, corresponding to an extremely disordered pattern. The exponent in the power varies as a function of parameters, i.e. is not universal (data not shown).

As the pulse solution and the extended endemic pattern can both occur at the same set of parameters, the emergence of the latter cannot be directly due to a linear instability of the former. Instead, there appears to be a finite amplitude instability wherein a nascent peak appears in the wake of a propagating pulse and manages to overcome the immune system inhibition. Once this occurs for the leading peak, the system cascades into the full extended pattern; this presumably occurs because the effective immune ``shadow” decreases with the number of patches as they compete for immune memory slots, as explained above for a two-pulse solution. This behavior gives rise to a rapid expansion phase, which only reaches the slow transient behavior described above when $N_I$ has reached its asymptotic value and the creation of new pulses becomes much less common.

It is of course quite difficult to devise analytic approaches to a complex multi-field nonlinear spatially-extended stochastic dynamical system. To make further progress, we next introduce a deterministic approximation to the dynamics, as already has been proven useful for the stable pulse solution in related models~\cite{levine-prl,rouzine, walczak1}. Our emphasis therefore will be on the extended pattern, whose existence and properties constitute the major new results of this work.

\section{Deterministic Equations}
\subsection{Derivation} 

We now proceed to derive a mean-field deterministic model that will allow us to gain some analytic insight into the co-evolutionary dynamics. We will work in the spatial continuum limit where differences and sums are replaced by derivatives and integrals. The equation for the density of infected hosts is just 
\begin{equation}
\dot{n}(x) \ = -\ r n(x) + I(x)
\label{ndot}
\end{equation}
where $r$ is the recovery rate and $I$ is the rate of new infections of type $x$.  The $I$ field in turn is given as the product of three factors, 
\begin{equation}
    I(x)=V(x) p_\textit{inf}(x) \left(1-\frac{N_I}{N_h}\right).
\end{equation}
Here, $V(x)$ is the rate of virion emission including the effects of mutation, while  $p_\textit{inf}(x) $ determines the immune inhibition.  The last factor accounts for the fact that  currently infected people cannot acquire a new infection.   In our stochastic model, the mean rate of virion emission by infected hosts of type $x$ is given by $r R_0 n(x)$.
 In constructing a deterministic model, we have also to incorporate the well-established idea~\cite{kepler,levine-dla,levine-prl,derrida,pechenik} that the production of new virions needs to be cut off at small $n$ as an approximate way to incorporate the discreteness of individual hosts. Thus the rate of emission by infected hosts carrying strain $x$ is taken to be
 \begin{equation}
 E(x) = r R_0 n(x) f_\epsilon(n(x))
 \label{Eeq}
 \end{equation}
 where $f_\epsilon$ is a function that goes from 0 to 1 as its argument increases and whose natural width depends on the cutoff parameter $\epsilon$.
 In particular, we choose the cutoff function 
\begin{equation} 
f_\epsilon (n) = 1 -e^{-(\frac{n}{\epsilon})^\alpha}
\label{feps}
\end{equation}
This function vanishes as $n^{\alpha}$ for $n \ll \epsilon$ and approaches 1 exponentially for $n >\epsilon$.

In our stochastic model, newly created virions can, with probability  $\mu$, represent a strain mutated with respect to their parent virus strain $x$, with their new phenotype being with equal probability either $x - 1$ or $x + 1$.  This translates to a total viral emission rate for strain $x$ of
\begin{equation}
    V(x)= E(x)  + \frac{\mu}{2}E''\label{Veq}
\end{equation}
 Note that in this continuum formulation, the only two quantities with dimensions of length are $\sqrt{\mu}$ and $w$, which were in the stochastic, lattice formulation, measured in units where the mutational change in $x$ was unity.  Thus, the relevant dimension-free parameter is the ratio $\mu/w^2$.

The remaining challenge is to devise a deterministic treatment of the immune memory and the concomitant construction of  $p_\textit{inf}$. We assume that there is no correlation between memories in the different slots of a given host or between hosts, so that the state of the memory is characterized simply by the memory density function $\rho_M (x,t)$. With this assumption, 
\begin{equation}
p_\textit{inf}(x) = \left( \int dx' \ \rho_M(x') \left(1-p_0 g (|x-x'|) \right)\right)^{M}
\label{pinf}
\end{equation}
where, as before. $g$ is an exponential with width $w$. Our stochastic simulations show that this is a quite good assumption, at least for $M=6$. Finally, we need to update the memory density. This occurs whenever a host recovers, and the resultant equation is
\begin{equation}
\dot{\rho}_M(x) = - \frac{\rho_M(x)}{M N_h} \int dx' \  r n(x') + \frac{rn(x)}{M N_h}  \ = \ 
- \frac{r \rho_M(x)N_I}{M N_h} + \frac{rn(x)}{M N_h}
\label{eq:rhomdot}
\end{equation}
The first term accounts for the overwriting of old memories, as each memory accounts for a fraction $1 / (MN_h)$ in the density. The second term refers to the writing of a new memory; note that the memory density remains normalized $\int \rho_M (x) dx = 1$, as the time derivative of this sum automatically vanishes. 

The fact that the infection probability is a non-local function of the memory density is an annoyance when it comes to both numerical calculation and analysis. It is therefore more convenient to define an auxiliary field $Q$ via
\begin{equation} Q'' - \frac{Q}{w^2} = -\frac{2}{w}  \rho_M ,
\end{equation}
supplemented by the boundary conditions that ensure that
\begin{equation}
Q(x) \ = \  \left( \int dx' \rho_M(x') g (|x-x'|) \right),
\label{Qint}
\end{equation}
in terms of which 
\begin{equation}
    p_\textit{inf}(x)=(1-p_0 Q(x))^M
\end{equation}
 $Q$ has no independent physical significance, as indicated by the fact that it has no actual dynamics; it is merely a way to track the factor by which infectivity is reduced at position $x$ by the integrated effect of all the current memories.  

We thus obtain the two dynamical equations in their final form
\begin{subequations}
\begin{align}
\dot{n}  &= -rn + I =  - rn + V (1-p_0 Q)^{M}\label{ndot1}\\
\dot{ \rho}_M &=  - \frac{r}{N_h M} \left(\rho_M N_I - n\right) \label{rhoMdot} 
\end{align}
\end{subequations}
with the auxiliary equations\begin{subequations}
\begin{gather}
E = r R_0 n \left(1-\frac{N_I}{N_h}\right)f_\epsilon (n) \label{Eeq1}\\
V = E + \frac{\mu}{2}E'' \label{Veq1}\\
Q'' - \frac{Q}{w^2} = -\frac{2}{w}\rho_M
\end{gather}
\end{subequations}
 Finally, an explicit global coupling in the model is through the dependence on $N_I$, the total number of infected individuals at a specific time, which enters into the above determination of the emitted virion field $E$. Of course, there is also global coupling through the $Q$ field.   

\subsection{Pulse solution} 
\begin{figure}[b]
\includegraphics[width=.5\linewidth]{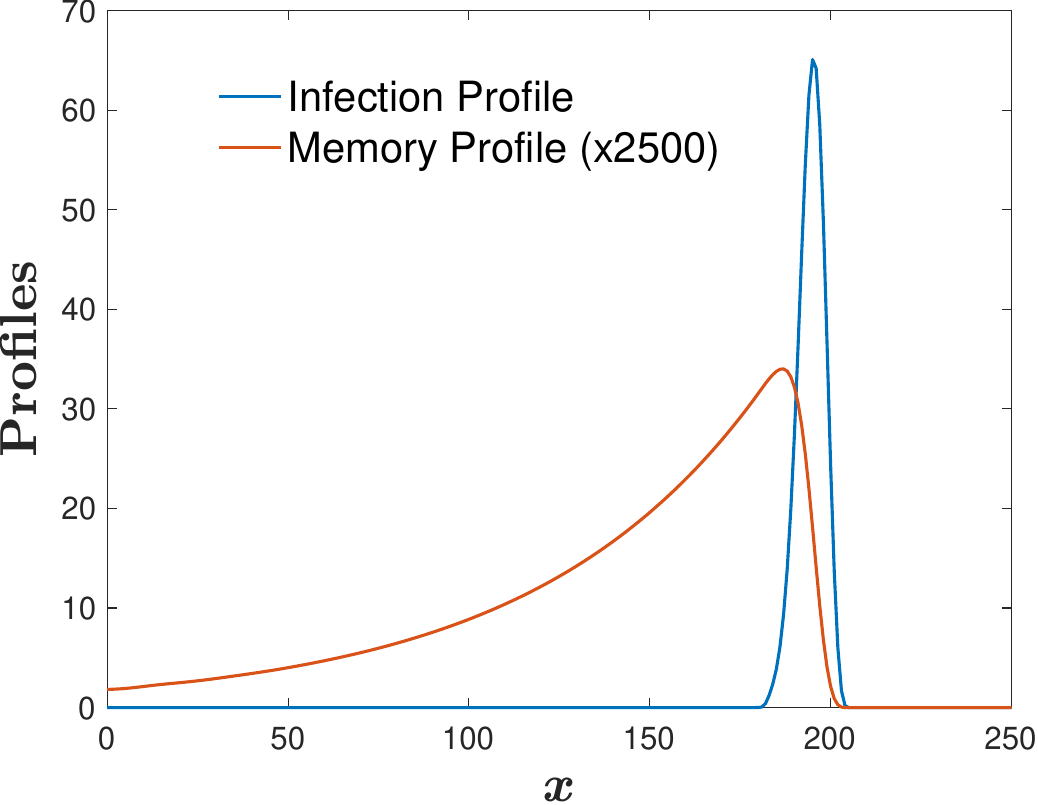}
\caption{Infection profile $n(x)$ and Memory Density (rescaled by a factor of 2500) $\rho_M(x)$ for a deterministic pulse, with $w=40$, $\mu=0.5$, $N_h=15000$, $\epsilon=1$, $\alpha=4$. \label{fig_5}}
\end{figure}
  We can now proceed to understand why the deterministic equation allows for an isolated pulse solution, moving with a steady-state shape at a specific constant velocity $c$ and containing a precise number of infected hosts. To see this, we need to analyze the deterministic equation in the two asymptotic regimes; the tail behind the front and the leading edge ahead of it. We can thereafter employ the idea of counting the number of parameters that need to be adjusted to satisfy matching conditions in order to determine the dimension of the space of allowed solutions.  As the derivation is fairly involved, we sketch it here and relegate the details to Appendix \ref{AppPulse}.

The lack of new infections in the tail implies that the 
 infected host density decays exponentially to the left (assuming a rightward propagation pulse), with an undetermined magnitude.  From Eq. \eqref{Eeq1}, we get that $E$ likewise decays exponentially, with a magnitude set by that of $n$. 
 From $E$ we can uniquely determine $V$ from Eq. \eqref{Veq1}. Also $\rho_m$ as given by Eq. \eqref{rhoMdot} has a homogeneous exponentially decaying solution, with an arbitrary amplitude as well as an inhomogeneous term driven by $n$. Lastly, $Q$ has a homogeneous term, with undetermined amplitude, that can be added to the particular solution driven by $\rho_m$.
   Thus the most general solution in the tail has three unknown constants,
   the amplitudes of $n$ and the homogeneous pieces of $\rho_m$ and $Q$.

The situation is considerably more complicated at the leading edge, as here we have new infections occurring. 
The cutoff in Eq. \eqref{ndot1} plays a key role here, as mentioned above. Since the cutoff not only multiplies the $E$ term in $V$, but also the ``diffusion" term $E''$, we effectively have a case of nonlinear diffusion. Hence we might expect (and will verify below) that solutions exist for which the infected host population becomes precisely zero after some value of $x$. This is analogous to what has been seen in the Fisher equation with nonlinear diffusivity ~\cite{nonlinear-fisher-1,nonlinear-fisher-2,eshel-review}; in particular, it has been shown there that the fronts that form from localized initial conditions have this specific property.  Thus, we have to first find the specific form of this nonlinear solution and then determine, by linearizing away from that solution, the number of unknowns governing the fields in this region. This calculation is presented in Appendix A, with the result that there is one unknown in this region, specifically $Q$ evaluated at the leading edge.

So, our two asymptotic solutions have a total of 4 unknowns, and adding in the velocity $c$ makes 5. This needs to be compared to the number of matching conditions if the equations are integrated separately from the two edges to some common point in the middle. $Q$ satisfies a second-order equation and hence both $Q$ and $Q'$ must be continuous. $\rho_M$ satisfies a first-order equation and hence it needs to be continuous. Since $I$ depends on two derivatives of $n$, the $n$ equation is effectively second-order; hence we must impose that $n$ and $n'$ are continuous. This gives a total of 5 equations, showing that in general there will be a discrete set of pulse solutions. Numerically, there appears to be only one stable solution of this type. A simulation starting from a localized source will indeed converge to such a solution. An example of a pulse solution is shown in Fig. \ref{fig_5}, where both the infected host population $n$ and the memory density $\rho_M$ are presented. We can explicitly check that this solution quantitatively agrees with the exponential behavior in the tail. The situation at the leading edge is trickier because of the singular behavior there. One must be careful to avoid effects due to the discrete nature of the phenotypic space (and of the simulation algorithm) in order to see the structure predicted by our spatially-continuous treatment. A simple approach yields qualitative but not extremely quantitative agreement. We will carry out this comparison in the next section, for a different type of solution.

As already mentioned, the single pulse solution in this model is similar in many respects to pulses found in other evolutionary dynamics problems. There are of course residual questions that are worth investigating, especially the dependence of the total infected number $N_I$ on the host population size $N_h$. However, the major point of this work is to understand the more exotic dynamical behavior found in the stochastic simulations, namely the extended patterns that emerge due to the ineffectiveness of the immune system in suppressing fluctuations that arise in the wake of a propagating pulse. This happens for large $N_h$, which translated in the deterministic model should correspond to small $\epsilon$. We now turn to a discussion of this behavior.
\begin{figure}[h]
\includegraphics[width=.32\linewidth]{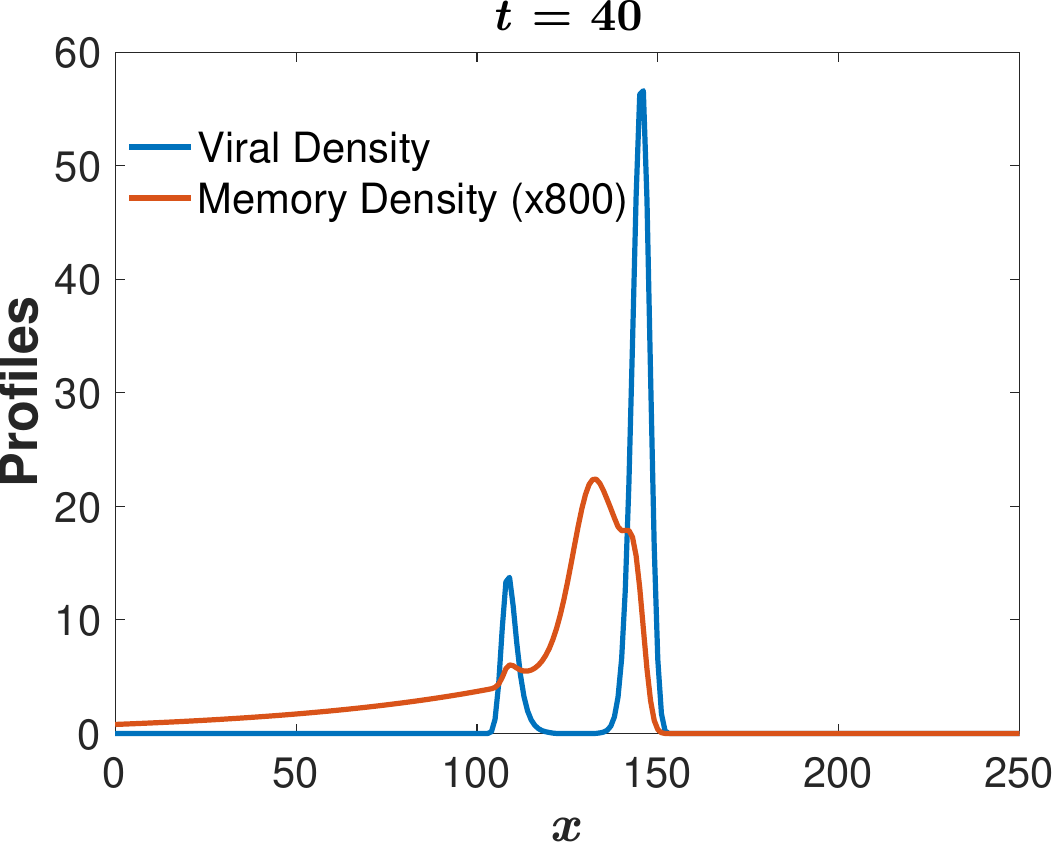}
\includegraphics[width=.32\linewidth]{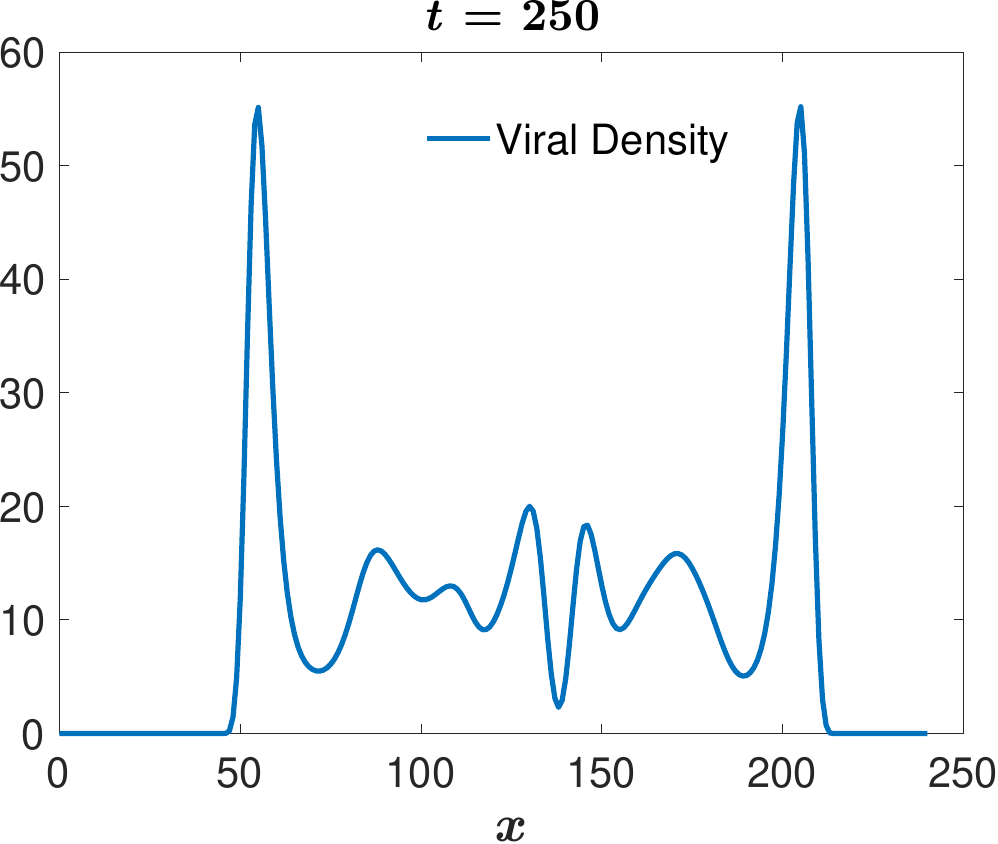}
\includegraphics[width=.32\linewidth]{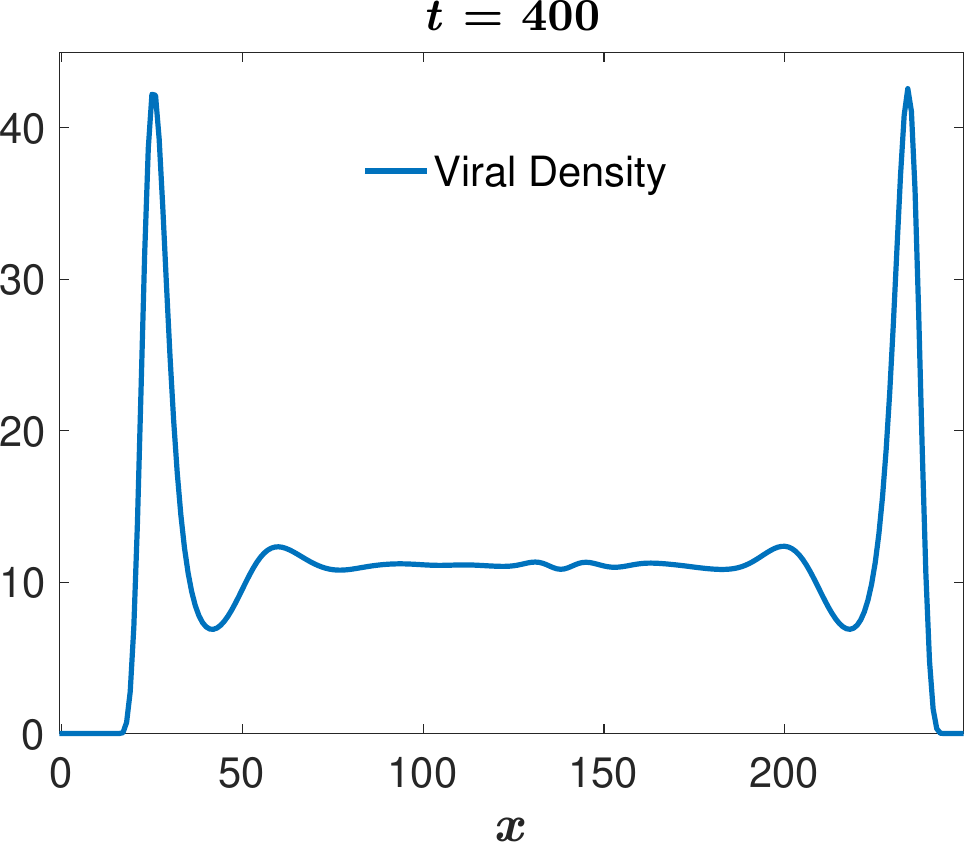}
\caption{Time development of a deterministic mound solution, for $w=20$, $\mu=0.125$, $N_h=5000$, $\epsilon=0.05$, $\alpha=4$. Left panel: Infection profile $n(x)$ (blue) and rescaled Memory profile $\rho_M(x)$ (red), at time $t=40$. Center panel: Infection profile $n(x)$ for $t=250$.\ Right panel: Infection profile $n(x)$ for $t=400$. \label{fig_6}}
\end{figure}

\subsection{Mounds}
In Fig. \ref{fig_6}, we present a series of snapshots from a simulation of the deterministic model at a low cutoff value. Initially a pulse is formed, but relatively quickly a second peak of the infected population density appears in the wake of the original pulse. This process can be seen in Fig. \ref{fig_6}a, where the second peak has appeared due to the failure of the immune ``shadow"  to adequately suppress its growth. This instability leads to a new type of static solution which we refer to as the mound. For these values of the parameters, this is a finite amplitude instability, as a careful preparation of the pulse state allows it to propagate indefinitely without this type of instability.

\begin{figure}[ht]
\includegraphics[width=.32\linewidth]{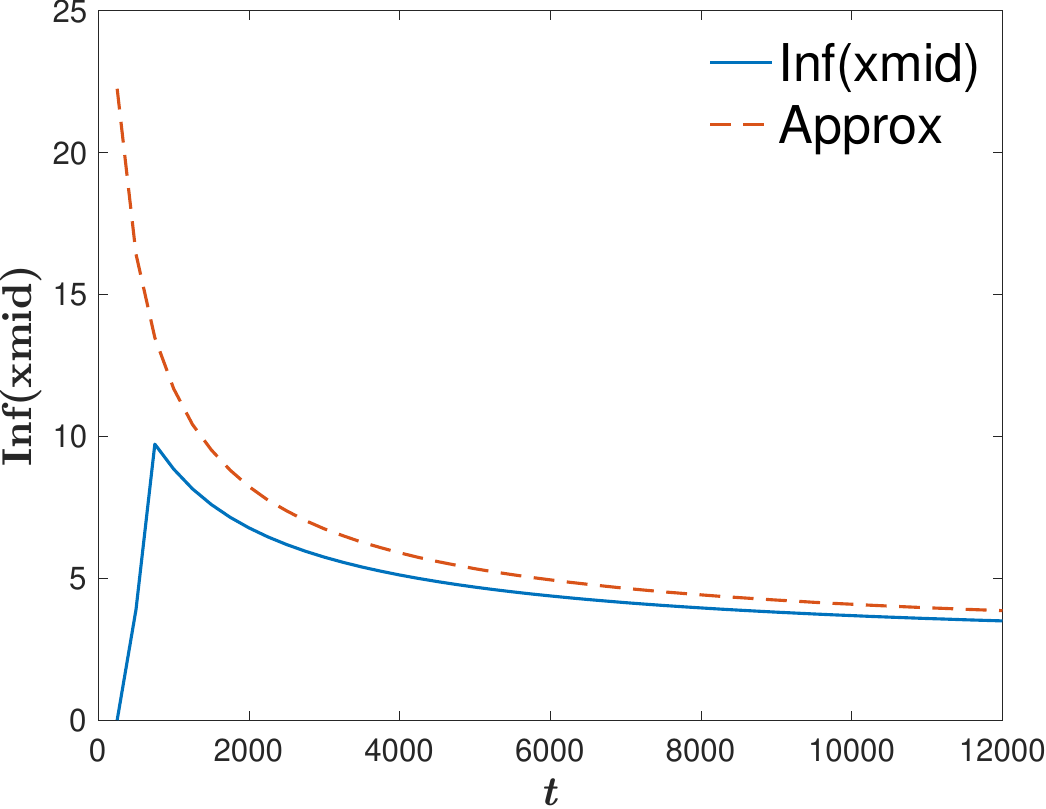}
\includegraphics[width=.32\linewidth]{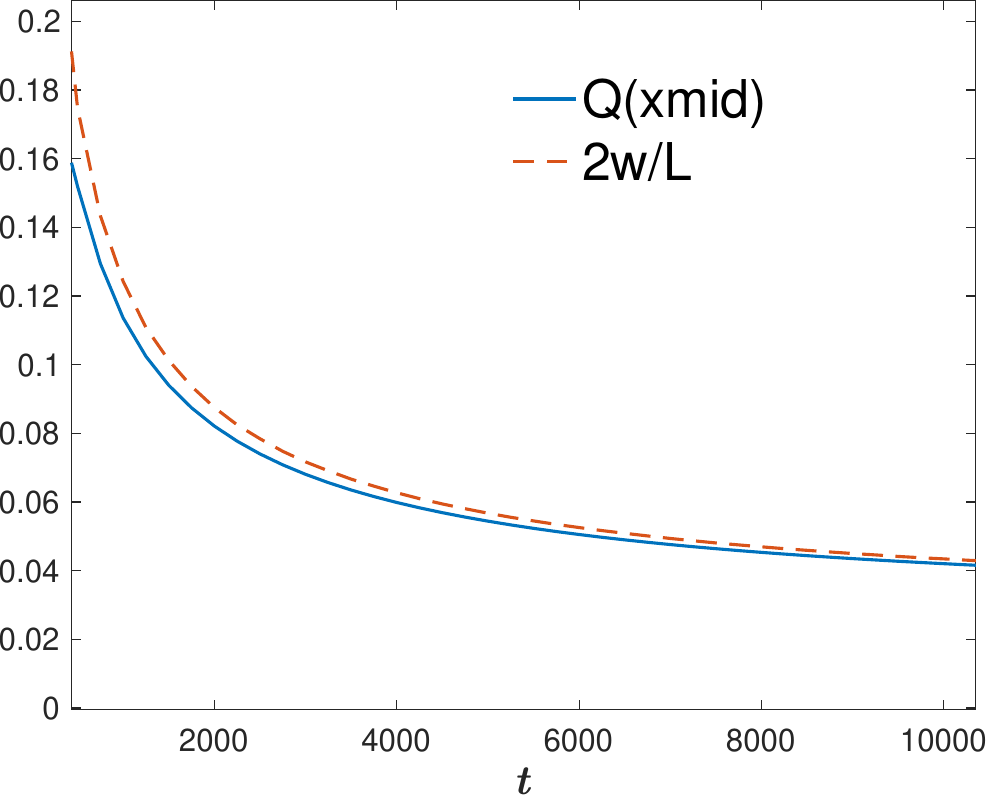}
\includegraphics[width=.32\linewidth]{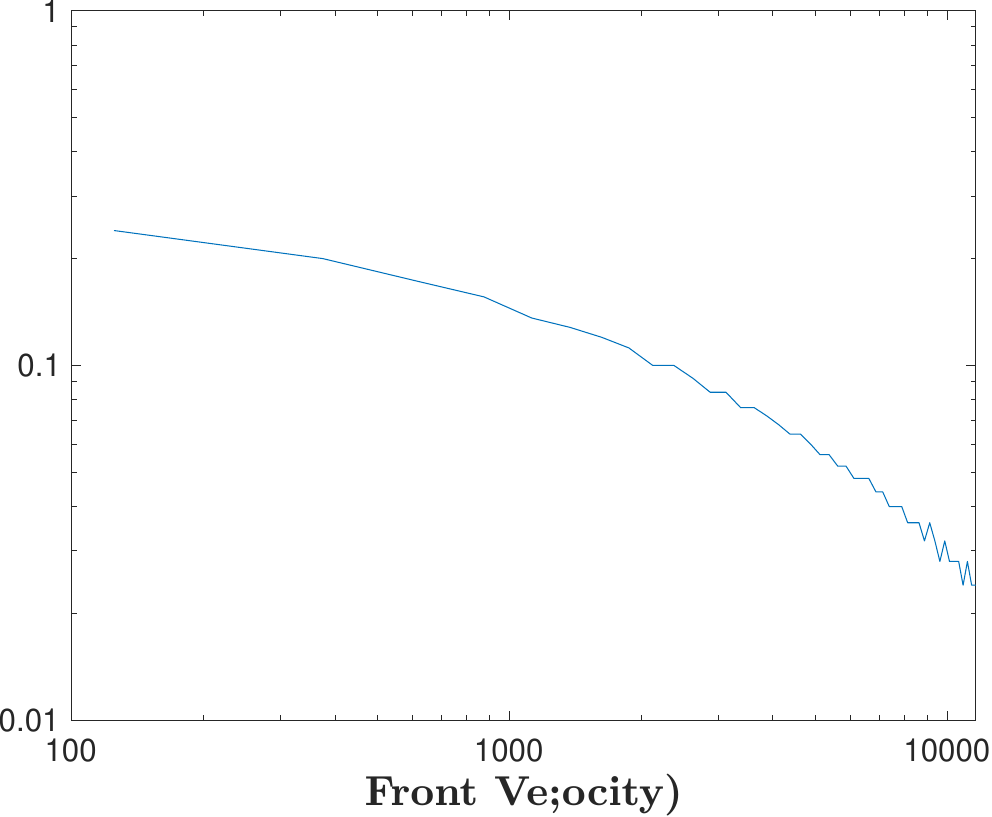}
\caption{ Comparison of numerical simulation with asymptotic theory from Eq. \ref{asymptotic}. The left block shows the infected population, the middle the auxiliary field $Q$, and the right is a plot of the decreasing velocity.} \label{expand}
\end{figure}

After the initial instability, the system settles into a slowly expanding pattern (Figs. \ref{fig_6}b and \ref{fig_6}c) with two counter-propagating peaks on the outside leaving behind a constant infected host density region between them. Let us assume that we can ignore for the moment the slow spreading of this constant region. Then, the bulk state obeys the time- and space-independent equations
\begin{align}
rn_0 &=   E (1-p_0 Q)^{M}\chi\nonumber\\
 \frac{Q}{w^2} & = \frac{2}{w}  \rho_M \nonumber \\
\rho_M &=  \frac{I(x)}{I_T} =  \frac{n_0}{N_I} \nonumber \\
E&= r R_0 n_0 f_\epsilon(n_0) 
\end{align}
The solution of this set of equations determines the bulk density $n_0$ via the implicit condition
\begin{equation}
\left(1-2w\frac{n_0}{N_I} p_0\right)^{M}=\frac{1}{R_0 \chi}
\end{equation}
This is valid in the case where $n_0$ is large enough such that the cutoff function can be taken to equal unity. Now, while the solution is expanding, we clearly have $N_I \approx n_0L$ so that $\chi\approx 1-n_0L/N_h$. Thus, the above equation gives rise to the simple predictions,
\begin{equation}
n_0 = \frac{N_h}{LR_0}  \left[R_0 - \left(1-\frac{2w}{L} p_0\right)^{-M} \right] \;\;\; ; \; \; Q_0 = \frac{2w}{L} \label{asymptotic}
\end{equation}
Once we have this result, we can qualitatively understand the spreading dynamics. The pair of pulses delineating the bulk regime will each move at a velocity $v$ that can be found from the type of two-pulse solution we have seen in the stochastic model, Fig. \ref{fig_3}b; with $\frac{dL}{dt} = 2v$. Here, however, the parameter $\chi$, which is related to the total number of infected hosts, and the global field $Q$ are no longer determined by the pulse solution itself, but instead are dominated by the bulk region. The reduction of $\chi$ as $N_I$ increases will slow the pulse, so that the velocity is time-varying. A comparison of these predictions with our numerical results is presented in Fig. \ref{expand}.

If there were no cutoff, the above argument suggests that the pulse would spread forever and would never reach a time-independent solution. This is of course not what we found for the stochastic model, where the system does eventually reach a statistical-steady state. The cause of this stopping is that eventually the cutoff becomes important in letting the pulse velocity go to zero. To understand this in more detail, we can consider the slightly simplified problem of an infinitely-long static "half-mound" which connects the aforementioned static solution as $x \rightarrow -\infty$ with a non-propagating pulse which completely vanishes at some particular $x$ value (again chosen to be $x=1$), as did the propagating pulse solution studied above. In this solution, the parameter $\chi$ is taken to be a free parameter and it directly determines $N_I$ through $N_I=N_h(1- \chi)$. Once the solution is found, however, it can be reinterpreted as a large but finite solution where the length $L$ is determined by the consistency requirement which is approximately $N_I = n_0 L$ or more exactly that the assumed value of $N_I$ is actual equal to $2\int_{1-L/2} ^1 n(x) dx $.

As with the pulse solution, we need to analyze separately the two asymptotic regimes, the semi-infinite bulk and the front, determining the degrees of freedom in both regions. We first consider the bulk region,  considering spatially dependent deviations from the constant bulk solution. Restoring the spatial derivatives and expanding around the solution, we have for the deviations of the various fields from their asymptotic limits,
\begin{align}
\delta_n &=   \delta_n \left(1+\frac{\mu}{2}k^2\right) - p_0 n_0 M (1-p_0 Q_0)^{-1}\delta_Q \nonumber\\
\delta_Q \left(k^2-\frac{1}{w^2}\right) &= -\frac{2}{wN_I}\delta_n
\end{align}
This yields the dispersion relation
\begin{equation}
(w^2 k^2-1)k^2 = -\frac{p_0 n_0 M w}{\mu N_I(1-p_0 Q_0)}
\end{equation}
In general, this gives rise to four complex modes of the form $e^{\pm ax \pm ibx}$.  Thus, as $x \rightarrow -\infty$, there are two bad and two good
modes, implying that the solution for large negative $x$  has two unknown parameters, namely the coefficients of the two good modes.

We must next turn to the leading edge. This is presented in detail in Appendix B. The result is the same as before, namely that the only new degree of freedom is the value of $Q$. We then are finally in a position to count degrees of freedom and compare to the continuity conditions that must be imposed to connect the two asymptotic regimes, say at $x=0$.  As compared to the propagating pulse, there is one fewer matching condition, namely four here, as the order of the $n$ equation has been reduced by 1. The number of unknowns is similarly four; two coefficients at large negative $x$, $Q(1)$ and $\chi$. Thus there is expected to be a unique solution, or perhaps a discrete set of solutions. Our simulations strongly suggest that there is only one such solution that is dynamically stable, as the system is always attracted to a unique mound state. 

\subsection{Shooting method}
We have implemented a shooting method to solve for the half-infinite mound. Based on the above analysis, we shoot forward from an arbitrarily chosen $x_b$, where we choose a small amplitude perturbation of the bulk solution.  The perturbation is characterized by two matching parameters, the phase of the growing fluctuation and $N_I$.  The fourth-order set of ODEs is integrated forward to the point where $n'(x)$ vanishes and $n$ is sufficiently large that we are out of the tail region.  In parallel, we shoot backward from the point $x=1$ where the density vanishes. Here, the matching parameters are $Q(1)$ and $N_I$. Again, we integrate till $n'(x)$ first vanishes. At this point, the other three fields have to agree, giving us three equations in three unknowns.  

\begin{figure}[h]
\includegraphics[width=.48\linewidth]{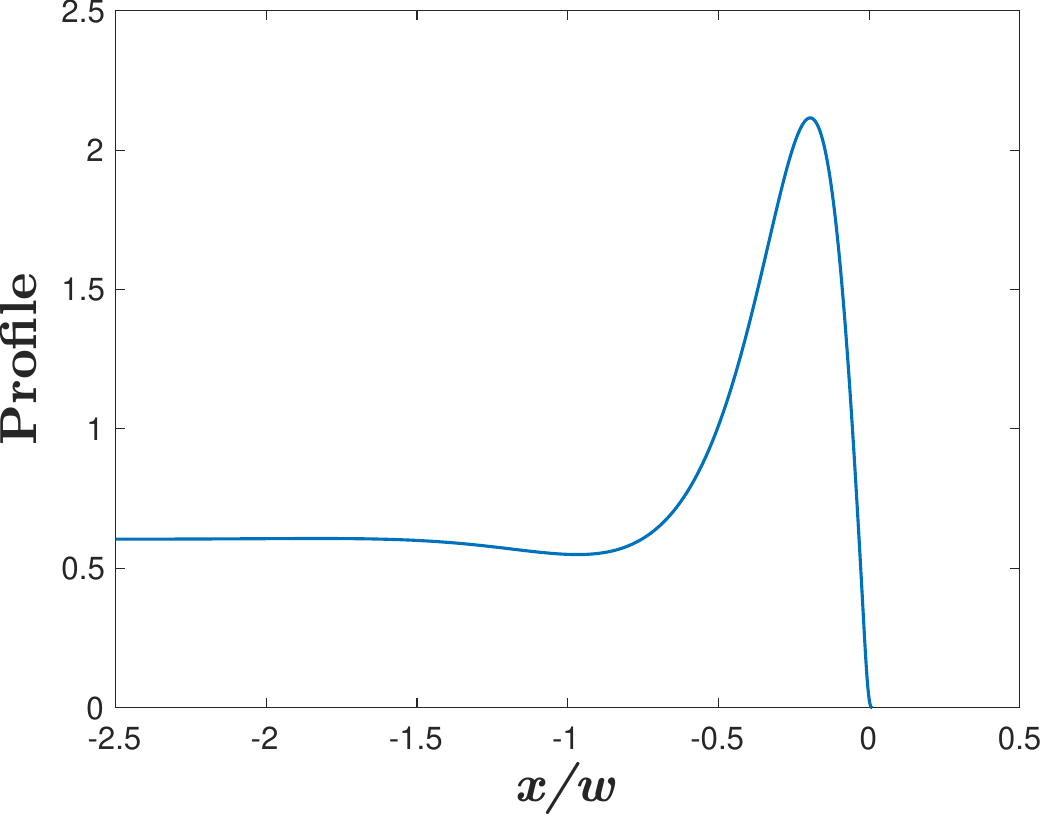}
\includegraphics[width=.48\linewidth]{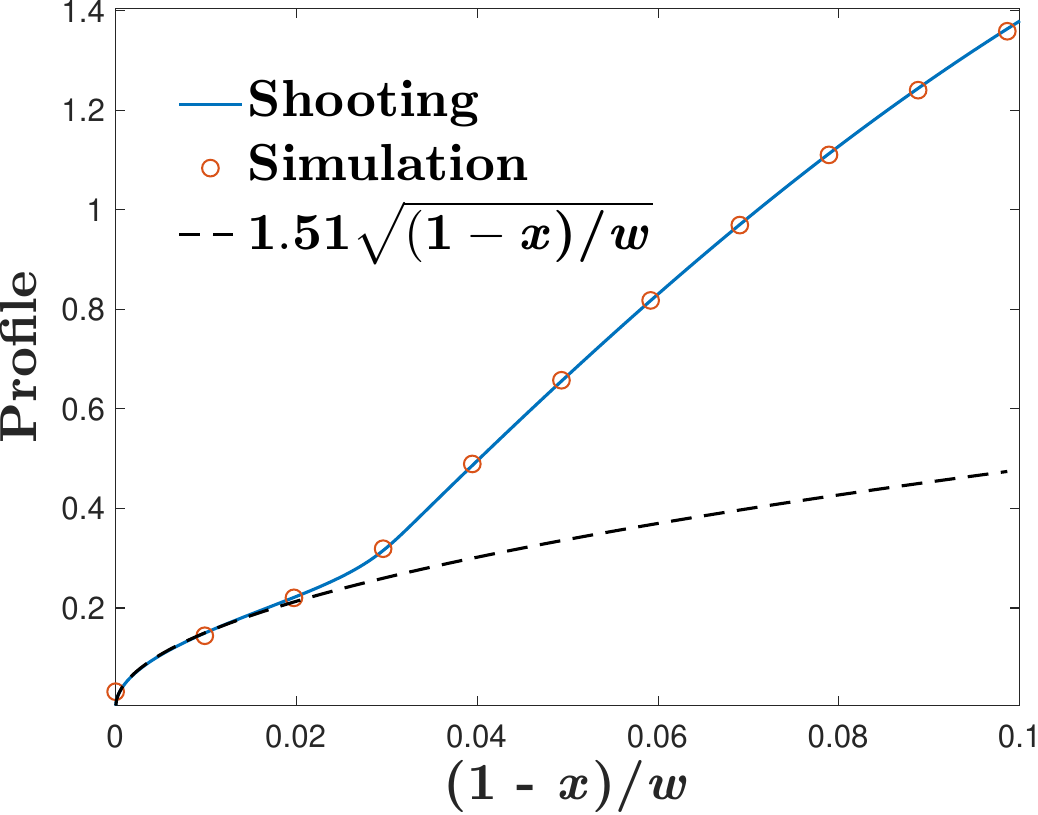}
\caption{Left panel: The half-infinite mound infection profile $n(x)$ as found by shooting for $\mu/w^2=2.5\cdot 10^{-4}$, $N_h=5000$, $\epsilon=0.22$, $\alpha=4$. Right panel: A blow-up of the leading edge of the half-mound solution, along with the tip region of the stationary mound obtained from a dynamical simulation on a discrete lattice and also the predicted $\sqrt{1-x}$ behavior. \label{fig:halfmound}}
\end{figure}

The resulting numerical solution is plotted in Fig. \ref{fig:halfmound} and matches exactly that produced by the time-dependent simulation. In particular, the behavior of the simulated system at the $x=1$ singular point exactly matches that predicted by the analytical treatment. To make this occur, we had to choose a very large value of $\mu$ so as to limit the effects of phenotypic discreteness (i.e. the spatial lattice with constant one) introduced by the finite difference simulation strategy.  At smaller values of $\mu$, the numerical solutions become affected by the lattice and the quantitative (but not qualitative) level of agreement degrades.

\subsection{Combs}
\begin{figure}[t]
\includegraphics[width=.5\linewidth]{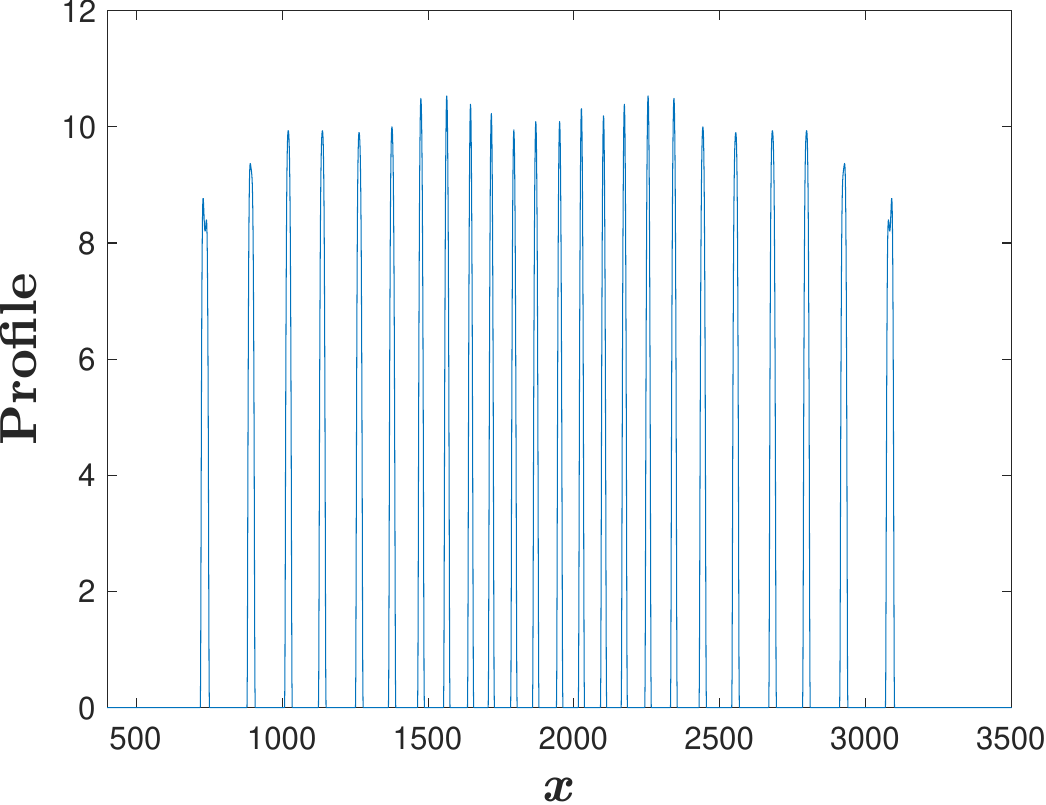}
\caption{The deterministic comb pattern infection profile $n(x)$ for $w=40$, $\mu=0.5$, $N_h=5500$, $\epsilon=1$, $\alpha=4$. \label{comb}}
\end{figure}
There is one additional pattern that is possible in the deterministic system. Fig. \ref{comb} shows what we will call a ``comb” configuration. This is a time-independent solution of the governing equations that again emerges via an instability of the tail of an originally propagating pulse. Here, however, the immune shadow behind each peak is large enough to prevent the full mound from forming, thereby leaving behind a pattern of isolated peaks.

Comparing the deterministic findings to the stochastic simulations, we see rough congruence.  Of course, the deterministic system can never exhibit a complete collapse of the infected population; this collapse is a fluctuation-induced effect that ultimately depends on the absorbing state nature of the $n=0$ solution. Aside from this, we saw that both models exhibit two types of states, a propagating local pulse versus an extended, eventually statistically steady, configuration. The difference between the deterministic mound and comb states has been obfuscated in the stochastic dynamics. The time average of the stochastic simulation is clearly consistent with the mound state even though a single snapshot is perhaps more analogous to a comb structure with its isolated peaks. For both cases, the number of infected hosts is a significant percentage of the total number of those present, unlike the case of the pulse solution with $N_I$ much less than $N_h$.

 \section{The role of memory decay \label{sec:memory}}

In the previous sections, we have demonstrated the existence of a tail instability that often leads to the formation of an endemic state of infection. This was shown initially in the full stochastic model and then analyzed in detail using a deterministic PDE approximation thereof. We note in passing that this instability was eliminated by hand in Ref. \cite{rouzine2} by choosing a highly non-symmetric infectivity profile which for a pulse never forgets previous infections. The physical justification for such a kernel is not clear and in any case does not appear to accurately describe respiratory infections (see below). A similar ad-hoc modification of the model was responsible for eliminating the instability in Ref. \cite{walczak2}. This instability arises from the decay of memories in our model due to the finite size of the memory buffer. We have checked that there can be no such instability if we allow an unbounded set of memories. We also checked that increasing the memory size will push the instability to larger values of $N_h$, assuming that all other parameters remain unchanged. In other words, there needs to be more available hosts to support the endemic state if each host has a more effective immune memory system.

The assumption that the new memory induced by the recovery from a new infection must replace an existing one is perhaps most closely realized in the CRISPR-based bacterial immune context~\cite{CRISPR-review}. There, virus information is stored in a DNA array which remains roughly at fixed size even as new viruses need to be remembered~\cite{crispr-decay}. There is still some uncertainty as to the precise molecular mechanism that enables this behavior and whether there is any bias in which memory is deleted to allow room for a new one to be stored. We have therefore checked that one finds an analogous tail instability (leading again to an endemic state) if one changes our model so as to drop the oldest rather than a randomly chosen memory. An example of the still-evolving endemic state along with the relevant infectivity profile is presented in Fig 10.  Going to this new model variant lowers the memory size threshold above which the tail instability vanishes. This occurs because dropping the oldest memory makes the immune system more potent to the variants currently dominant in the system.

What about the human adaptive immune system and other similar ones? There is significant evidence that at least for some respiratory viruses such as the SARS family, there is indeed a decay of immune response over extended times~\cite{sars-decay,covid-decay1}. This decay apparently involves the reduction of the number of long-lived antibodies, as has been shown for SARS directly and more recently for COVID-19~\cite{covid-decay2}. It does not appear to be true for other viral diseases such as measles. Thus, diseases with memory loss typically lead to the SEIRS class of models~\cite{seirs}, corresponding to susceptible, exposed, infected, recovered, and returning to susceptible subpopulations. Our model is of course a variant-resolved version of SEIRS, albeit with a more complex return to susceptibility depending (in our case) on the degree of mutations.
\begin{figure}[t]
\includegraphics[width=.8\linewidth]{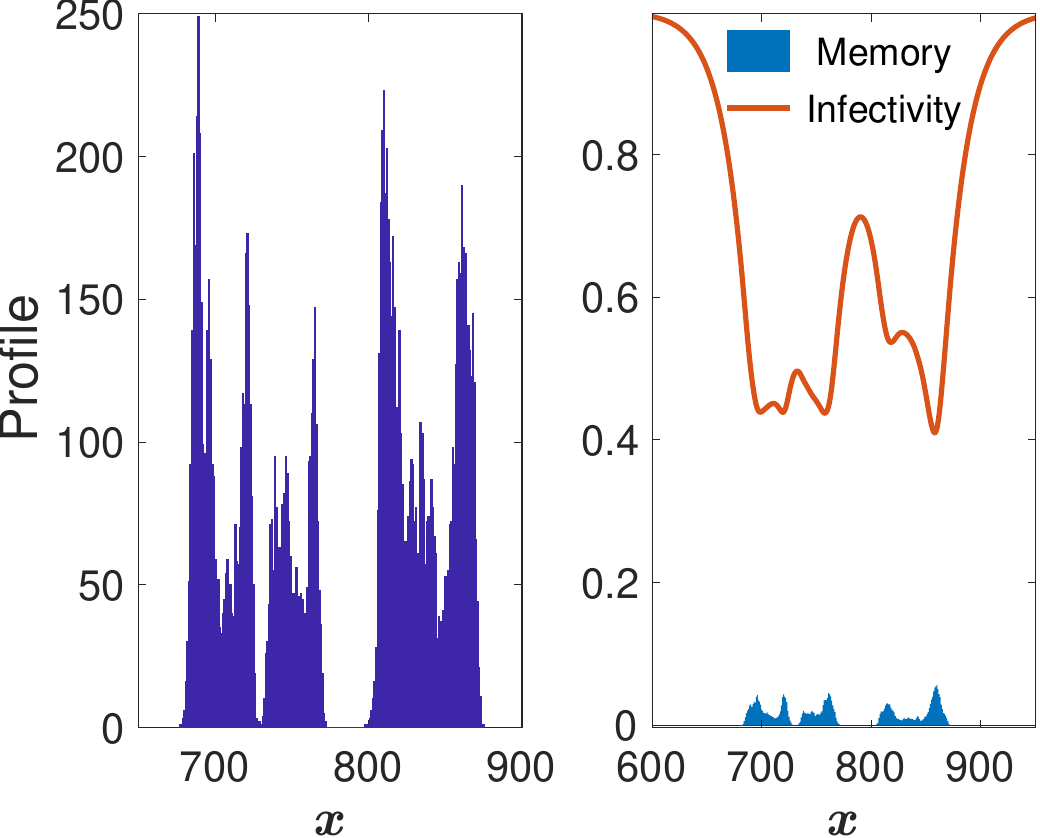}
\caption{  Simulation of a case where the oldest memory is replaced upon new infection. Here, a snapshot of the nascent mound state is shown, for $N_h = 30K$, $w=20$, $\mu =.5$ and a memory buffer of size 3. \label{memory-variant}}
\end{figure}

To test whether the precise form used above, involving memory replacement as opposed to simple loss, is essential, we can instead introduce memory decay as a simple stochastic process. Here, we eliminate memory replacement and in its place introduce a rate at which an individual memory decays. This additional stochastic process becomes part of the overall Gillespie algorithm, which now chooses appropriately between individual host recovery, new infection of a host, and decay of a randomly chosen memory. In Fig. 11, we present simulation snapshots from this new variant model, first at the ratio of memory decay to host recovery of 0.005 (showing no tail instability) and then at 0.03 (showing a tail instability).  These results clearly indicate that the precise form of memory decay is not an essential aspect of the new endemic state. We can therefore expect that both respiratory viruses in mammals as well as phages in bacteria might exhibit this possibility.
\begin{figure}[h]
\includegraphics[width=.48\linewidth]{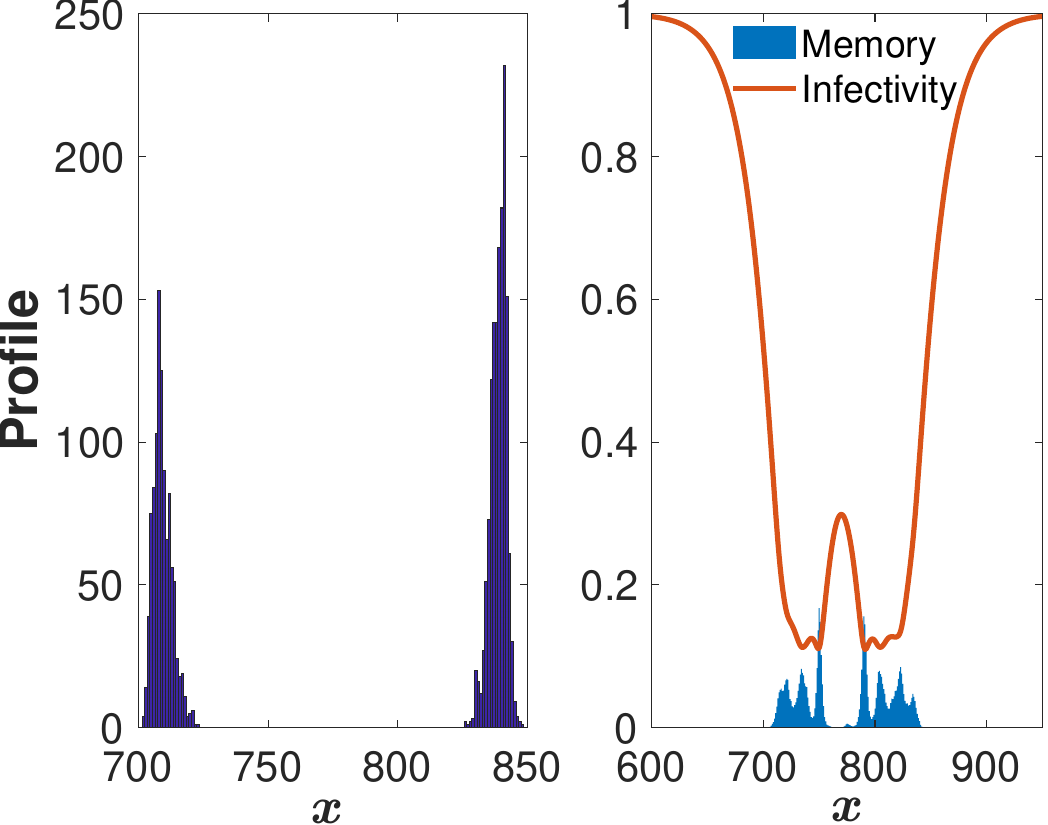}
\includegraphics[width=.48\linewidth]
{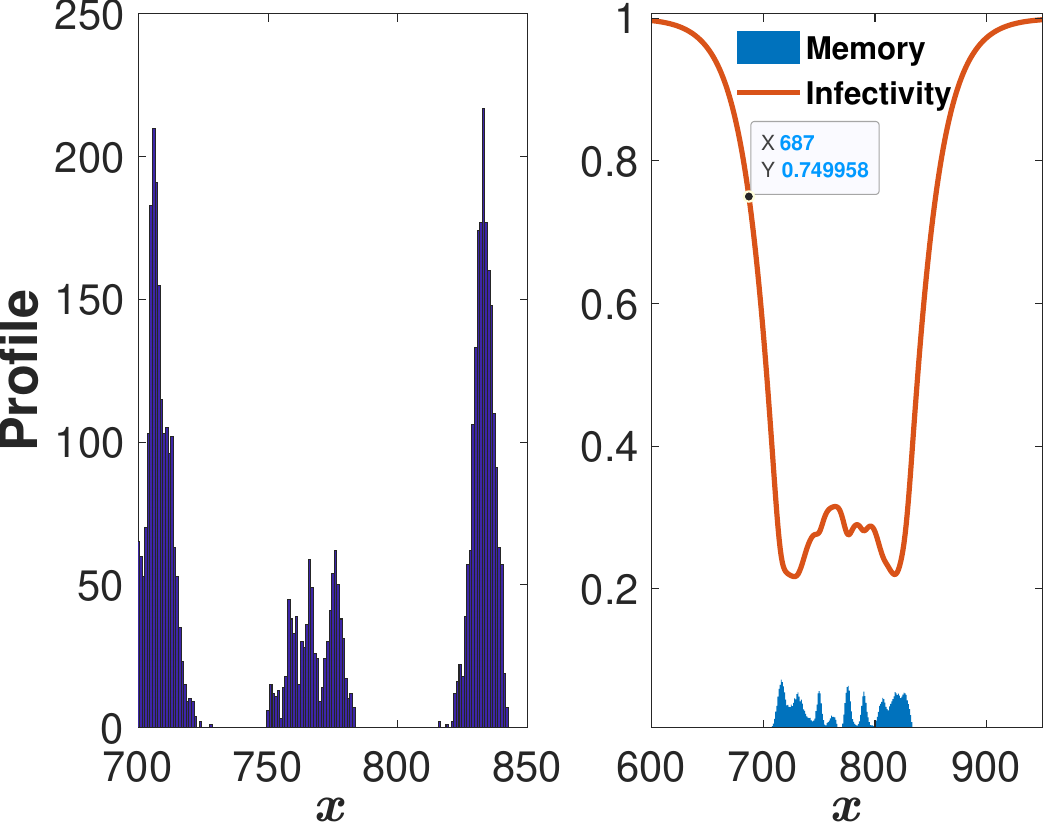}
\caption{ Simulations of the memory decay variant where there is no a priori limit on memory size, but individual memories decay with variable rates. Other parameters are the same as in Fig. 11.  In the left two panels, the ratio of memory decay rate to infection recovery rate is 0.005, and in the right two panels it is 0.03. Note the difference in infectivity for the two different cases. \label{memory-variant-2}}
\end{figure}

For this set of parameters, a ratio of 0.01 is at the margin, which means that some runs result in extended states and some do not. Interestingly, this value seems to be in the relevant physiological range for COVID-19~\cite{covid-decay2}, several years versus a week or two.  Basically, the issue is the extent to which individuals left behind the main pulse remain infected long enough for their viral loads to take advantage of memory decay in the population and ignite a new burst. This new burst must rely on increased infectivity and if we compare the two subfigures of Fig. 11, we see the difference in the infectivity (roughly 0.3 versus 0.1) in the unstable versus stable cases. We also point out that we study the case of two counter-propagating pulses due to convenience, as this is the state that is easiest to produce starting from a localized infected host population as an initial condition. All of our findings equally apply to one propagating pulse, albeit with some change (i.e. lowering) in the stability threshold due to the elimination of the requirement for hosts to remember two separate co-existing strains.

\section{Discussion}

Understanding the never-ending battle between viruses and our adaptive immune system is clearly a critical issue in modern-day statistical physics and biomedicine. The problem and its importance have been amply demonstrated by our experience with COVID-19.  A new virus emerges, and the immune response is constantly challenged by the ability of the virus to mutate to new variant strains that are beyond the range of immune coverage. We certainly need to do a better job of understanding the resultant dynamics and devising possible ways to mitigate the corresponding consequences.

Viral evolution has been effectively modeled by devising equations that govern population propagation in a fitness landscape. These ideas have been successfully applied both to artificial evolution in laboratory experiments as well as to HIV infections. However, most of these works have assumed that the aforementioned fitness landscape, however complex, was not able to adapt to changing viral populations. This is explicitly not the case when considering the role of the immune system in responding to viral infection. Thus, several recent papers on coupling viral evolution to dynamic immune system response represent a welcome expansion of the applicability of fitness evolution modeling.

This work has presented a new formulation of viral-immune co-evolution,  including the effects of a finite host population and the dynamics of a (possibly) finite immune memory capacity. We have shown that in addition to the pulse-type solution discussed at length in previous works, there are other radically different possible behaviors. Specifically, there can be an extended state in which the host population remains infected by a very broad range of viral strains. This state initially appears via a finite-amplitude instability of the pulse solution and leads to a pattern of endemic infection whose spread is only limited by the total host population and the ensuing demographic noise. We have studied this novel pattern both by direct simulation and also by a semi-analytic approach relying on a cutoff mean-field deterministic version of the governing stochastic process. Our model contains, of course, many parameters, but our results suggest that a key determinant of the observed behavior is the range of immune coverage, here characterized by the ratio of the width $w$  of the exponential fall-off in immune efficacy to the phenotypic distance covered by viral evolution during a typical host infection time,  here $c/r$ where $c$ is the pulse speed.  Once the system reaches an extended configuration, its eventual size is determined by the effective cutoff, which here scales as $1/N_h$.  

We have already discussed how adaptive systems with effective memory decay should generically exhibit the phenomenology described in this work. We wish to add that such a decay can occur at the individual level as in the variety of models considered above, each exploring a different specific mechanism, or could occur also at the population level. For example, imagine that we are considering populations with birth and death, the latter being necessary to address questions regarding disease virulence and its possible evolution. Then, new individuals would constitute a non-immunized reservoir which could drive our tail instability. Another possibility is an infection that is spreading geographically, reaching previously uninfected hosts. Taken all together, we expect that many real-life situations could lead to the type of endemic states discussed here. 

 It is important to note that the tail instability provides a temporal pattern distinct from what would be seen in a system exhibiting the type of oscillatory state found by Sasaki and co-workers~\cite{japanese}. There, the pulse exhibits a “near-death” episode and only survives when small numbers of individuals at the leading edge start growing rapidly. In this scenario, a new strain would appear only after a period of time with no obvious infection; based on our preliminary investigations, this behavior is in fact very sensitive to the form of the cross-reactivity function and is not at all seen in models with the exponential decay used here (data not shown; results to be published). In our case, the basic infection continues to evolve continuously but spawns reinfection by strains that naively appeared to be vanquished.

There are many additional directions that could be studied in future work. The reduction of the problem to a one-dimensional phenotypic equation relies on the localization of the viral population in orthogonal dimensions, namely neutral genomic variations that do not contribute to fitness changes. This localization has been seen in simulations of the pulse solution (\cite{walczak1} and unpublished data) but has not to date been considered in the context of extended states. From the pure theory perspective, proving that the pulse instability is always finite-amplitude is a tractable problem. There is also the need to better understand the $N_h$ dependence of the various patterns, possibly finding an explicit way to take the limit $N_h \rightarrow \infty$ for the pulse solution. It should also be possible to relate the ``comb" solution to similar structures that can appear in the type of non-local Fisher equations used in the modeling of species diversity~\cite{pigolotti,rogers,kessler}.

The real challenge for this line of research is making more quantitative contact with actual experimental data.  This will require a detailed analysis of the cross-reactivity function. In the context of COVID-19, there is data and analysis regarding how antibodies effective against one strain lose effectiveness when confronted with other strains \cite{dixit}. In the context of CRISPR-based phage immunity, there is a reasonable understanding of how changes in the viral sequence can alter the immune response~\cite{CRISPR-review}. Once this is accomplished, one should be able to look for the temporal signature of the tail instability.  One might imagine comparing the spreading of the mound solution (before it reaches a steady state) with data on the dynamics of viral variants in a host population. A pulse solution corresponds to a replacement theory; a new strain takes over and no one gets infected by previous strains. A mound state on the other hand suggests the possibility that old variants can re-appear even as old variants remain active. On the applied side, one could use our theoretical framework to design vaccine treatments that would interfere with the natural dynamics and increase the chances of collapse of the infected population. 
   
\begin{acknowledgments} HL acknowledges the support of the NSF of the Center for Theoretical Biological Physics,  PHY-2019745.  DAK acknowledges the support of the Israel Science Foundation, 1614/21. We thank Jonathan Levine for help with the schematic of the model.
\end{acknowledgments}

\bibstyle{apsrev4-2}
\bibliography{walczak}

\appendix
\section{Pulse Solution: Details\label{AppPulse}} 
In this appendix, we present more of the details of the argument arguing for at most a discrete set of propagation
velocities for the isolated pulse solution, each with a specific number of infected hosts. As outlined in the main text, the essence of the argument is to 
 analyze the number of adjustable parameters characterizing the solution of the deterministic equation in each of the two asymptotic regimes; the tail behind the front and the leading edge ahead of it. 

Starting with the tail, we immediately conclude that the lack of new infections there means that the infected host density decays exponentially. Specifically, replacing $\dot{n}$ by $-cn'$ (i.e.  going to the moving frame and assuming the pulse is moving to the right) and dropping $I$, we find from Eq. \eqref{ndot} that $ n \approx n_b e^{rx/c}$ where $n_b$ is an as yet unknown constant. Next we substitute this form into the definition of $E$ in Eq. \eqref{Eeq1}; using the aforementioned form of the cutoff we arrive at
\begin{equation}
E \ \approx  r R_0  \epsilon ^{-\alpha} e^{ -(\alpha +1) rx/c } n_b^{\alpha+1}.
\end{equation}
 From $E$ we can uniquely determine $V$ from Eq. \eqref{Veq1}. Finally we note that $\rho_m$ has a homogeneous solution and $Q$ has a homogeneous term that can be added to the particular solution driven by $\rho_m$,
\begin{eqnarray}
\rho_m  &\approx & \rho_{m,b} e^{\beta x} - \frac{n_b}{N_h M} e^{rx/c} \nonumber \\
Q &\approx & \frac{2\rho_{m,b}}{1/w - w \beta ^2 } e^{\beta x} + Q_b e^{x/w} 
\end{eqnarray}
where we have defined $\beta = N_I/(c MN_h)$.  The inhomogeneous term is exponentially supressed relative to the inhomogeneous term, since $\beta<r/c$. Thus the most general solution in the tail has three unknown constants, $n_b$, $\rho_{M,b}$ and $Q_b$.

The situation is considerably more complicated at the leading edge, as here we have new infections occurring. 
The cutoff in Eq. \eqref{ndot1} plays a key role here, as mentioned above. Since the cutoff not only multiplies the $E$ term in $V$, but also the ``diffusion" term $E''$, we effectively have a case of nonlinear diffusion. Hence we might expect (and will verify below) that solutions exist for which the infected host population becomes precisely zero after some value of $x$. This is analogous to what has been seen in the Fisher equation with nonlinear diffusivity ~\cite{nonlinear-fisher-1,nonlinear-fisher-2,eshel-review}; in particular, it has been shown there that the fronts that form from localized initial conditions have this specific property.  Thus, we have to first find the specific form of this nonlinear solution and then determine, by linearizing away from that solution, the number of unknowns governing the fields in this region. 

To accomplish this, we assume that
\begin{equation}
  n(x) \approx n_f (1-x)^\gamma,  
\end{equation}
where we have used translation invariance to fix the singular point to occur at $x=1$. The left-hand side of the $n$ evolution equation, Eq. \eqref{ndot1} is then $c n_f \gamma (1-x)^{\gamma -1}$.
This term is larger than the $-rn$ term on the right hand side and therefore it must be balanced by the infection piece $I$. As $x$ approaches 1 from below, the leading term in $I$ arises from the $E''$ term in $V$. Using the asymptotic form of the cutoff for small $n$, we have
\begin{eqnarray}
E &\approx& rR_0 
\epsilon^{-\alpha} n^{\alpha +1}   \approx rR_0 \epsilon^{-\alpha} n_f ^{\alpha +1} (1-x)^{(\alpha+1) \gamma }    \nonumber \\
V &\approx& \frac{\mu}{2\epsilon^{\alpha}} rR_0  n_f ^{\alpha +1} (\alpha+1)\gamma\left[(\alpha+1)\gamma -2\right]   (1-x)^{(\alpha+1) \gamma -2} 
\end{eqnarray} 
To get $I$, we just have to multiply by the as yet undetermined factor $\chi\equiv 1 - \frac{ N_I}{N_h}$ and by $p_\textit{inf}(x)\approx p_\textit{inf}(1)=(1- p_0 Q(1))^M$, since $Q$ is determined globally and is not singular. Thus, the matching of the two sides of Eq. \eqref{ndot1} requires the powers of $(1-x)$ to be equal to each other, which yields $\gamma = 1/\alpha$.  Then the coefficients must also match, fixing $n_f$ to satisfy
\begin{equation}
n_f^{-\alpha} \ = \ \frac{\mu r R_0 \chi (1+\alpha) }{2c \epsilon^\alpha \alpha} (1-p_0 Q(1))^M
\end{equation} This result shows that $Q(1)$ is an unknown number, adding to the list of parameters that need to be determined. Note that the integral formula for $Q$, Eq. \eqref{Qint} directly determines that
$Q'(1) = -Q(1)/w$ and hence this is not an additional degree of freedom.

We now consider the linearized equation around this base solution so as to determine how many unknowns  need to be specified in order to integrate the equations from the leading edge to the pulse center.
Denoting the shifted values of $n$, $E$ and  $I$ as respectively $\delta n = A_n (1-x)^p$, $\delta E = A_E (1-x) ^{p+1}$ and $\delta I = A_I (1-x) ^{p-1}$, we have the following equations to leading order in $(1-x)$;
\begin{eqnarray}
cp A_n & = & A_I \nonumber \\
A_I &=& (1-p_0 Q(1))^M \frac{\mu}{2} p (p+1) A_E \nonumber \\
A_E &=& rR_0 n_f^\alpha (\alpha+1) A_n \epsilon^{-\alpha} \ = \ A_n \frac{2c\alpha}{\mu (1-p_0Q(1))^M} \nonumber
\end{eqnarray}
Combining these equations leads to the very simple condition, $p = p(p+1) \alpha$, which yields the two possibilities $p=0$ and $p = 1/\alpha -1$. The second of these just corresponds to a translation of the base solution (recall that $\gamma = 1/\alpha$) and neither of these are allowed. Thus there are {\emph{no}} additional unknown constants in the leading edge solution. We also should mention that $\chi$ is an unknown parameter, but this will have to be determined self-consistently by the fact that it is directly related to $N_I$; picking any arbitrary value of $\chi$ will lead to this constraint being violated.

These results are used in the main text to show that the number of unknown parameters (including the velocity as one of these) equals the number of matching conditions. Thus, constructing a global solution will fix the velocity, leading to there being at most a discrete set of allowed pulses. In fact, we never find more than one such solution.

\section{Mound Solution: Details\label{AppMound}}
In the main text, we considered the dimension of the solution space emerging when considering spatially dependent deviations from the constant bulk solution. We showed that there are two bad and two good
modes, implying that the solution for large negative $x$  has two unknown parameters, namely the coefficients of the two good modes.

We here turn to the front edge. Unlike what was the case for the propagating front, the time-independent equation for $n$ is purely algebraic and hence $n$ can be formally eliminated, leaving us with a two-field problem in terms of $E$ and $Q$. In detail, 
we will eliminate $n$ via Eq. \ref{Eeq1}; since the right-hand side is a monotonically increasing function of $n$, we can define $g(x)$ such that 
\begin{equation}
    n(E) = g(E/r R_0)
\end{equation}
so that Eq. \eqref{Eeq1} translates to
\begin{equation}  g(\zeta)(1-e^{-(g(\zeta)/\epsilon)^\alpha})=\zeta
\end{equation}
Then the coupled differential equations read
\begin{align}
E+ \frac{\mu}{2}E '' &= \frac{rn(E)}{(1-p_0 Q(x))^{M}\chi}\nonumber\\
Q'' - \frac{Q}{w^2} &= -\frac{2}{w}  \frac{n(E)}{N_I} \nonumber
\end{align}
 We now have to count degrees of freedom, and to do this we have to investigate the behavior of $E$ near
 $x=1$. Writing $E \approx E_1 (1-x)^\gamma$,
 we first need the small argument expansion of $g$:
 \begin{equation}
 g(\zeta)^{\alpha+1}/\epsilon^\alpha \approx \zeta \qquad\Rightarrow\qquad  g(\zeta) \approx \epsilon^{\alpha/(\alpha+1)} \zeta^{1/(\alpha+1)}
 \end{equation}
 so that 
 \begin{equation}
 \frac{\mu}{2}\gamma(\gamma-1)E_1(1-x)^{\gamma-2} \approx r \epsilon^{\alpha/(\alpha+1)} \left(\frac{E_1(1-x)^\gamma}{rR_0}\right)^{1/(\alpha+1)}\frac{1}{(1-p_0 Q(1))^{M}\chi}
 \end{equation}
 From this, we can read off $\gamma$:
 \begin{equation}
 \gamma-2 = \frac{\gamma}{\alpha+1} \Rightarrow  \gamma = \frac{2(\alpha+1)}{\alpha}
 \end{equation}
 This implies that $n \sim (1-x)^{\gamma-2} = (1-x)^{2/\alpha}$.  Solving now for $E_1$ yields
 \begin{equation}
 E_1^{\alpha/(\alpha+1)} = \frac{2 r}{\mu\gamma(\gamma-1)} \epsilon^{\alpha/(\alpha+1)} \left(\frac{1}{rR_0\chi}\right)^{1/(\alpha+1)}\frac{1}{(1-p_\textit{imm}^0 Q(1))^{N_M}}
 \end{equation}
 so that
 \begin{equation}
 E_1 = \left(\frac{\alpha^2}{\mu(\alpha+1)(\alpha+2)(1-p_0 Q(1))^{M}\chi} \right)^{(\alpha+1)/\alpha} r\epsilon
 \left(\frac{1}{R_0}\right)^{1/\alpha}
\end{equation}
 This implies that
 \begin{equation}
 n\approx  \left(\frac{\alpha^2}{\mu(\alpha+1)(\alpha+2)(1-p_0 Q(1))^{M}R_0\DeclareGraphicsRule{}{}{}{}} \right)^{1/\alpha} \epsilon
 (1-x)^{2/\alpha}
 \end{equation}
 
As for the pulse,  we now need to investigate the behavior of the linearized equations about this nonlinear solution.  Writing
 \begin{equation}
E  \approx E_1(1-x)^\gamma + \delta_E; \qquad Q \approx Q(1) + \delta_Q
 \end{equation}
 we have the linearized system (to leading order in $(1-x)$)
 \begin{align}
 \frac{\mu}{2}\delta_E'' &= \delta_E \frac{rn'(E)}{(1-p_0 Q(1))^{M}}=
r \delta_E  \frac{n(E)}{E(1+\alpha)(1-p_0 Q(1))^{M}} \nonumber\\
&\approx
\delta_E  \frac{\mu(E'')}{2E(1+\alpha)}  = \delta_E  \frac{\mu(\alpha+2)}{\alpha^2(1-x)^2} \nonumber\\
\delta_Q'' &= -\frac{2}{w} \delta_E \frac{n'(E)}{N_I} 
= -\frac{2}{w}\frac{n(E)}{(1+\alpha)EN_I}\delta_E = -\frac{\mu E'}{rwE(1+\alpha)N_I}\delta_E\nonumber\\
&\approx-\frac{2\mu(\alpha+2)}{rw(1-x)^2N_I}\delta_E
\end{align}
The assumption $\delta_E \sim (1-x)^p$ is consistent with the spatial dependence of these two equations. Substituting this into the first equation, we get the indicial equation for the exponent $p$,
\begin{equation}
p(p-1)=\frac{2(\alpha+2)}{\alpha^2} \Rightarrow  p=\{ \frac{\alpha+2}{\alpha} , -2\alpha\}
\end{equation}
The first of these choices, being $\gamma-1$, corresponds to translations (i.e. moving the location of the singular point), and so is ruled out.  The second is singular and is also forbidden, so in fact the solution for $E$ is unique.  $Q$ satisfies the boundary condition
$Q'(1)=-(1/w) Q(1)$, so the only degree of freedom at the edge is $Q(1)$.

\end{document}